\documentclass[twoside, twocolumn, 9pt]{article}
\usepackage[utf8]{inputenc}
\usepackage[T1]{fontenc}
\usepackage{graphicx}
\usepackage{longtable}
\usepackage{wrapfig}
\usepackage{rotating}
\usepackage[normalem]{ulem}
\usepackage{amsmath}
\usepackage{amssymb}
\usepackage{float}
\usepackage{colortbl}
\usepackage{capt-of}
\usepackage{hyperref}
\usepackage[acronym,nogroupskip,nonumberlist]{glossaries}
\usepackage[stylemods]{glossaries-extra}
\makeglossaries
\usepackage{abstract}
\usepackage[export]{adjustbox}
\usepackage{sectsty}
\usepackage{setspace}
\usepackage{amsthm}
\usepackage{mathtools}
\usepackage{lastpage}
\usepackage{amsfonts}
\usepackage{mathptmx}
\usepackage{charter}
\usepackage{times}
\usepackage{bm}
\usepackage[
format=plain,
justification=justified,
singlelinecheck=false,
font={small, sf},
labelfont=bf,
labelsep=space
]{caption}

\usepackage[compact]{titlesec}
\usepackage{fancyhdr}
\usepackage{fnpos}
\usepackage{sidecap}
\usepackage{gensymb}
\usepackage{upgreek}
\usepackage{soul}

\usepackage{array}
\usepackage{droidsans}

\usepackage[usenames,dvipsnames]{xcolor}
\usepackage[version=3]{mhchem}
\usepackage[super,sort&compress,comma]{natbib}

\usepackage[
left=1.5cm,
right=1.5cm,
top=1.785cm,
bottom=2.0cm
]{geometry}

\usepackage{orcidlink}
\usepackage{microtype}

\definecolor{cream}{RGB}{222,217,201}

\newacronym{dft}{DFT}{density functional theory}
\newacronym{sqs}{SQS}{special quasi-random structures}
\newacronym{paw}{PAW}{projector augmented wave}
\newacronym{gga}{GGA}{generalized gradient approximation}
\newacronym{pbe}{PBE}{Perdew-Burke-Ernzerhof}
\newacronym{hse}{HSE}{Heyd-Scuseria-Ernzerhof}
\newacronym{soc}{SOC}{Spin Orbit Coupling}
\newacronym{ma}{MA}{Methylammonium}
\newacronym{fa}{FA}{Formamidinium}
\newacronym{pce}{PCE}{power conversion efficiency}
\newacronym{hap}{HaPs}{halide perovskites}
\newacronym{pca}{PCA}{principal component analysis}
\newacronym{tsne}{t-SNE}{t-distributed stochastic neighbor embedding}
\newacronym{umap}{UMAP}{uniform manifold approximation and projection}
\newglossaryentry{vasp}{name=VASP,description={{Vienna Ab initio Simulation Package}}}
\newglossaryentry{slme}{name=SLME,description={{spectroscopic limited maximum efficiency}}}
\newglossaryentry{qmml}{name=QM/ML,description={{quantum mechanics machine learning}}}
\date{\today}

\begin{document}


\pagestyle{fancy}
\thispagestyle{plain}
\fancypagestyle{plain}{
\renewcommand{\headrulewidth}{0pt}
}

\makeFNbottom
\makeatletter
\renewcommand\LARGE{\@setfontsize\LARGE{15pt}{17}}
\renewcommand\Large{\@setfontsize\Large{10pt}{14}}
\renewcommand\large{\@setfontsize\large{10pt}{12}}
\renewcommand\footnotesize{\@setfontsize\footnotesize{7pt}{10}}
\makeatother

\renewcommand{\thefootnote}{\fnsymbol{footnote}}
\renewcommand\footnoterule{\vspace*{1pt}%
\color{cream}\hrule width 3.5in height 0.4pt \color{black}\vspace*{5pt}} 
\setcounter{secnumdepth}{5}

\makeatletter 
\renewcommand\@biblabel[1]{#1}            
\renewcommand\@makefntext[1]{\noindent\makebox[0pt][r]{\@thefnmark\,}#1}
\makeatother 
\renewcommand{\figurename}{\small{Fig.}~}
\sectionfont{\sffamily\Large}
\subsectionfont{\normalsize}
\subsubsectionfont{\bf}
\setstretch{1.125}
\setlength{\skip\footins}{0.8cm}
\setlength{\footnotesep}{0.25cm}
\setlength{\jot}{10pt}
\titlespacing*{\section}{0pt}{4pt}{4pt}
\titlespacing*{\subsection}{0pt}{15pt}{1pt}

\fancyfoot{}
\fancyfoot[RO]{\footnotesize{\sffamily{1--\pageref{LastPage} ~\textbar  \hspace{2pt}\thepage}}}
\fancyfoot[LE]{\footnotesize{\sffamily{\thepage~\textbar\hspace{3.45cm} 1--\pageref{LastPage}}}}
\fancyhead{}
\renewcommand{\headrulewidth}{0pt} 
\renewcommand{\footrulewidth}{0pt}
\setlength{\arrayrulewidth}{1pt}
\setlength{\columnsep}{6.5mm}
\setlength\bibsep{1pt}

\makeatletter 
\newlength{\figrulesep} 
\setlength{\figrulesep}{0.5\textfloatsep} 

\newcommand{\topfigrule}{\vspace*{-1pt}%
\noindent{\color{cream}\rule[-\figrulesep]{\columnwidth}{1.5pt}} }

\newcommand{\botfigrule}{\vspace*{-2pt}%
\noindent{\color{cream}\rule[\figrulesep]{\columnwidth}{1.5pt}} }

\newcommand{\dblfigrule}{\vspace*{-1pt}%
\noindent{\color{cream}\rule[-\figrulesep]{\textwidth}{1.5pt}} }

\makeatother

\twocolumn[
\begin{@twocolumnfalse}

\vspace{1em}
\sffamily
\begin{tabular}{m{2.5cm} p{14.5cm} }
& \noindent\LARGE{\textbf{Accelerating Defect Predictions in Semiconductors Using Graph Neural Networks}}\\%
\vspace{0.3cm} & \vspace{0.3cm} \\
& \noindent\large{Md. Habibur Rahman\textsuperscript{a}, Prince Gollapalli\textsuperscript{b,c}, Panayotis Manganaris\textsuperscript{a}, Satyesh Kumar Yadav\textsuperscript{b,c}, Ghanshyam Pilania\textsuperscript{d}, Brian DeCost\textsuperscript{e}\,\orcidlink{0000-0002-3459-5888}, Kamal Choudhary\textsuperscript{e}, and Arun Mannodi-Kanakkithodi\textsuperscript{a*}\,\orcidlink{0000-0003-0780-1583}}\\
\end{tabular}


\begin{abstract}

First principles computations reliably predict the energetics of point defects in semiconductors, but are constrained by the expense of using large supercells and advanced levels of theory. Machine learning models trained on computational data, especially ones that sufficiently encode defect coordination environments, can be used to accelerate defect predictions. Here, we develop a framework for the prediction and screening of native defects and functional impurities in a chemical space of Group IV, III-V, and II-VI zinc blende (ZB) semiconductors, powered by crystal Graph-based Neural Networks (GNNs) trained on high-throughput density functional theory (DFT) data. Using an innovative approach of sampling partially optimized defect configurations from DFT calculations, we generate one of the largest computational defect datasets to date, containing many types of vacancies, self-interstitials, anti-site substitutions, impurity interstitials and substitutions, as well as some defect complexes. We applied three types of established GNN techniques, namely Crystal Graph Convolutional Neural Network (CGCNN), Materials Graph Network (MEGNET), and Atomistic Line Graph Neural Network (ALIGNN), to rigorously train models for predicting defect formation energy (DFE) in multiple charge states and chemical potential conditions. We find that ALIGNN yields the best DFE predictions with root mean square errors around 0.3 eV, which represents a prediction accuracy of 98~\% given the range of values within the dataset, improving significantly on the state-of-the-art. Models are tested for different defect types as well as for defect charge transition levels. We further show that GNN-based defective structure optimization can take us close to DFT-optimized geometries at a fraction of the cost of full DFT. DFT-GNN models enable prediction and screening across thousands of hypothetical defects based on both unoptimized and partially-optimized defective structures, helping identify electronically active defects in technologically-important semiconductors.

\end{abstract}
\end{@twocolumnfalse}
\vspace{0.6cm}
]

\renewcommand*\rmdefault{bch}\normalfont\upshape
\rmfamily
\section*{}
\vspace{-1cm}


\footnotetext{%
  \textsuperscript{a} School of Materials Engineering, Purdue University, West Lafayette, IN 47907, USA; E-mail: amannodi@purdue.edu

  \textsuperscript{b} Department of Metallurgical and Materials Engineering, Indian Institute of Technology (IIT) Madras, Chennai 600036, India
  
  \textsuperscript{c} Centre for Atomistic Modelling and Materials Design, Indian Institute of Technology (IIT) Madras, Chennai, 600036, India
  
  \textsuperscript{d} GE Research, Schenectady, NY, 12309, USA
  
  \textsuperscript{e} Materials Measurement Laboratory, National Institute of Standards and Technology, Gaithersburg, MD, 20899, USA

}
\footnotetext{%
  \textsuperscript{\dag}Electronic Supplementary Information (ESI) available:
  \url{https://github.com/msehabibur/defect_GNN_gen_1}
  See DOI: 00.0000/00000000.
}


\section{Introduction}
Semiconductors are critical for a variety of technologies such as consumer electronics, healthcare and biotechnology, information technology, communication and connectivity, automotive manufacturing, renewable energy, and industrial automation\cite{Ganose_Scanlon_2022}. With the signing of the CHIPS Act \cite{CHIPS}, there has been a massive influx of funding into semiconductor R\&D, resulting in the establishment of several manufacturing facilities and research centers across the United States as well as many global partnerships between companies and universities. Developing next-generation semiconductor materials is crucial for addressing global energy needs and the demands of the electronics industry, and this process begins at the atomistic scale with enhanced structure-property relationships that can scale up to device performance and aid data-driven materials design and improvement\cite{J_2022}. \\

The electronic structure of a semiconductor is heavily dependent on the presence of point defects in its crystal lattice, which range from intrinsic vacancies, self-interstitials, and anti-site substitutions, to impurities at different lattice sites\cite{Chen_Walsh_Gong_Wei_2013}. Defects can introduce energy levels within the band gap, affecting carrier concentration and mobility, and often acting as traps that lead to non-radiative recombination of holes and electrons in optoelectronic devices\cite{Mannodi-Kanakkithodi_2023, Defect_Energetics_JPCC, Chemistry_of_MSE, Broberg_Bystrom, Buckeridge_2019, Turiansky}. Defects also play a crucial role in dopant activation, diffusion, and segregation, which are vital factors in device fabrication processes. Even at low concentrations, unwanted point defects or impurities can have a significant impact on the electronic, optical, and transport properties of semiconductors, making it important to be able to accurately predict their stability and electronic signatures \cite{Lannoo1981}. \\

One of the applications where the effect of defects is felt most is solar cells, where semiconductors such as Si and CdTe are commonly used as absorbers\cite{Mannodi-Kanakkithodi2022-ck, Chemistry_of_MSE}. Undesirable defects and functional dopants in semiconductor absorbers will respectively decrease and increase the optical absorption and thus the power conversion efficiency of single-junction, tandem, and bifacial solar cells\cite{Chen_Yang_2010}. Similar effects are felt in applications such as transistors, photodiodes, lasers, sensors, and quantum information sciences\cite{Queisser_Haller_1998, Liu_2020, Nayak_2019, Sivathanu_R_Lenka_2021, Kazeev_Al}. Canonical group IV, III-V, and II-VI semiconductors are some of the most important materials used in these applications, either as binary compounds or alloys, typically in the zinblende (ZB) or wurtzite (WZ) phases. In addition to Si and CdTe, compounds such as GaAs, SiC, CdSe, and CdS have been used in photovoltaics (PVs). GaAs, GaN, ZnO, and InP are employed in optoelectronic devices such as light-emitting diodes (LEDs), laser diodes, quantum dots, and quantum wells. GaN, AlN, and related compounds are desirable wide band gap (WBG) semiconductors for power electronics \cite{Gorai}. Point defects negatively or positively impact the performance of each of these materials; in general, semiconductors with intrinsic defect tolerance and possible n-type or p-type dopability are desired for optoelectronic applications. Furthermore, defect levels in semiconductors (e.g., NV-centers in diamond) have also been suggested as qubits for quantum computing \cite{nv} . \\

Experimentally, defect levels are measured using techniques such as cathodoluminescence and deep-level transient spectroscopy\cite{Srivastava_Ranjan}. However, these methods face difficulties in sample preparation and assigning measured levels to particular defects; e.g., it is not trivial to determine whether a captured peak is from a specific vacancy or self-interstitial, or from an unwanted substitutional or interstitial impurity\cite{Mannodi-Kanakkithodi2022-ck}. First principles-based density functional theory (DFT) computations have thus been widely used to calculate the formation energy (E$^{f}$) of point defects as a function of Fermi level (E$_{F}$), net charge in the system (\textit{q}), and chemical potential ($\mu$)\cite{Kim_Park_Hood_Walsh_2019, Rahman_Mannodi-Kanakkithodi_2023}. Such computations help reliably identify the lowest energy donor and acceptor defects, all possible shallow and deep defect levels, the equilibrium conductivity as pinned by the lowest energy defects (p-type, intrinsic, or n-type), defect concentrations, electron/hole capture rates, and other related properties. When using an appropriate level of theory, DFT-computed charge transition levels have been shown to match remarkably well with measured levels \cite{Mannodi-Kanakkithodi2022-ck}. Despite the successes of DFT, large-supercell charged calculations are rather computationally expensive, which makes it prohibitive to perform defect calculations across a very large chemical space. \\ 

Predicting defect properties can be significantly accelerated by combining DFT data with state-of-the-art machine learning (ML) models\cite{Mannodi_Toriyama_2020, Mannodi-Kanakkithodi2022-ck}. Some recent efforts, including our own past work, have shown that regression models trained on a DFT dataset, using vectors encoding the identity and elemental properties of the coordinating atoms involved in creating a defect, can yield accurate prediction and screening over tens of thousands of defects and impurities \cite{Witman_2023,def_ml1,def_ml2}. In published work from 2022 \cite{Mannodi-Kanakkithodi2022-ck}, we trained ML models to predict the formation energies and charge transition levels of point defects in 34 ZB semiconductors with a roughly 90\% prediction accuracy, which enabled qualitatively reliable screening of consequential impurities from across a list of $>$ 12,000. However, these models suffer from the following limitations: (a) for a wide chemical space, composition-based models \cite{Bartel} require a significant amount of training data to accurately capture the complex relationships between the material and target properties, (b) all predictions are made for a supposed ground state defect structure which means no competing defective structures could be sampled and no lower energy defect structures could theoretically be found, (c) the errors are larger than desired, presumably due to lack of information in the model inputs about the defective geometry and how local coordination changes in the presence of different defects, and (d) the predictive models cannot trivially be applied to related but ``out-of-sample" defects, such as complexes and alloys within the same material classes. A potential approach to tackle these issues arises in the form of a ``crystal graph", which is the most general representation of a crystalline structure, automatically accounting for different supercell sizes, types of ionic species, mixing or added atoms, and metastable configurations. \\

\begin{figure*}[!ht]
\centering
\includegraphics[width=.9\linewidth]{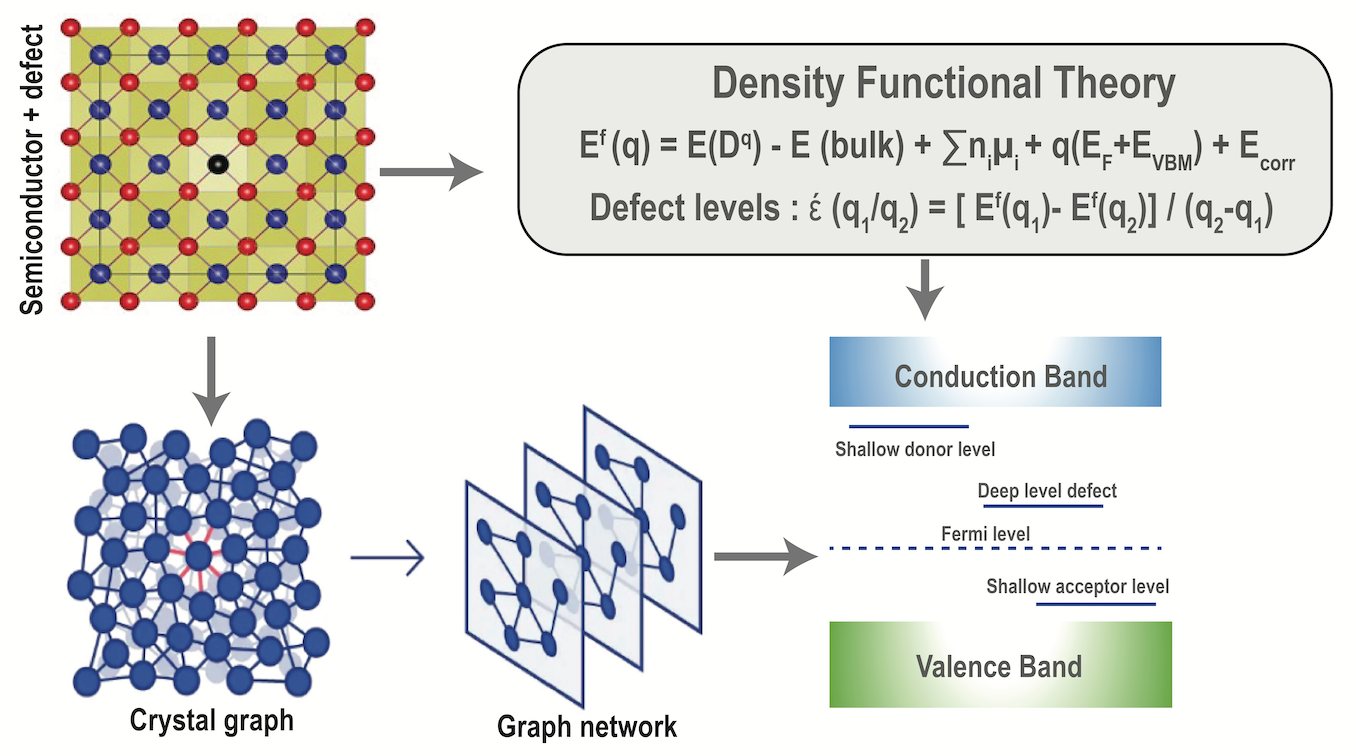}
\caption{\label{fig:outline} DFT+GNN workflow to accelerate the prediction of defect formation energies and charge transition levels in semiconductors.}
\end{figure*}

Graph-based Neural Networks (GNNs) have been rising in popularity over the last few years, and are widely applied today for adequately representing and predicting properties of molecules, polymers, and inorganic crystals\cite{Fung, Chen_Ong_2022, Cheng, Park, Bhattacharya_2023, Witman_2023, Kilgour_Rogal_Tuckerman_2023, Bapst_Keck}. GNNs can work directly with graph-structured data, converting crystal structures into crystal graphs where the nodes are atomic positions and edges are chemical bonds. They are capable of learning internal representations of crystals useful for predicting properties ranging from formation or decomposition energy to band gap to defect formation energy. ML models based only on vectors encoding composition and/or elemental properties are typically not suited to deal with crystalline polymorphs of a given material, often requiring hand-crafted features that are not generalizable. GNNs are known to be much more flexible than composition-based models, as they can be normalized with respect to system size or number of atoms, and have the ability to capture important structure/chemistry information that contribute to properties of interest. By learning the global representation of crystals, GNNs can incorporate physical laws and phenomena on larger scales and be used to predict properties that are affected by long-range interactions. Predictive models trained using GNNs show much better accuracy than models that lack structural/geometry information. \\

In this article, we present one of the most comprehensive efforts undertaken to date for predicting defect properties in semiconductors, by combining a massive high-throughput (HT) DFT dataset of defect formation energies (DFEs) with state-of-the-art GNNs. We utilize our previously published dataset \cite{Mannodi-Kanakkithodi2022-ck, Mannodi_Toriyama_2020}, bolstered by the inclusion of thousands of partially-optimized and unoptimized structures in addition to optimized structures, along with several new computations, and train predictive models using three types of established GNN schemes: Crystal Graph Convolutional Neural Network (CGCNN) \cite{CGCNN}, Materials Graph Network (MEGNET) \cite{MEGNET}, and Atomistic Line Graph Neural Network (ALIGNN) \cite{ALIGNN}. We train GNN models on datasets ranging from a few thousand to nearly 15,000 data points for point defects in different charge states, across 40 or so unique ZB compounds. We present a description of model optimization and visualization of the prediction results for different defect types and show how (a) ALIGNN predictions significantly improve upon previous DFE estimates, with a root mean square error (RMSE) of $\sim$ 0.3 eV, (b) predictions can be made for defect complexes and alloyed compositions by including a subset of them in the training data, and (c) GNN predictions for new systems can be used both for screening based on upper-bound energies as well as for stabilizing any defect by changing the defect coordination environment until the GNN-predicted formation energy minimizes. We believe this provides a novel and promising approach towards predicting defect energetics and screening important defects from large chemical spaces. Considerations of the level of theory and future prospects of this work are additionally discussed. The DFT+GNN workflow applied for predicting defect properties is presented in \textbf{Fig. \ref{fig:outline}}. \\

\begin{figure*}[t]
\centering
\includegraphics[width=.9\linewidth]{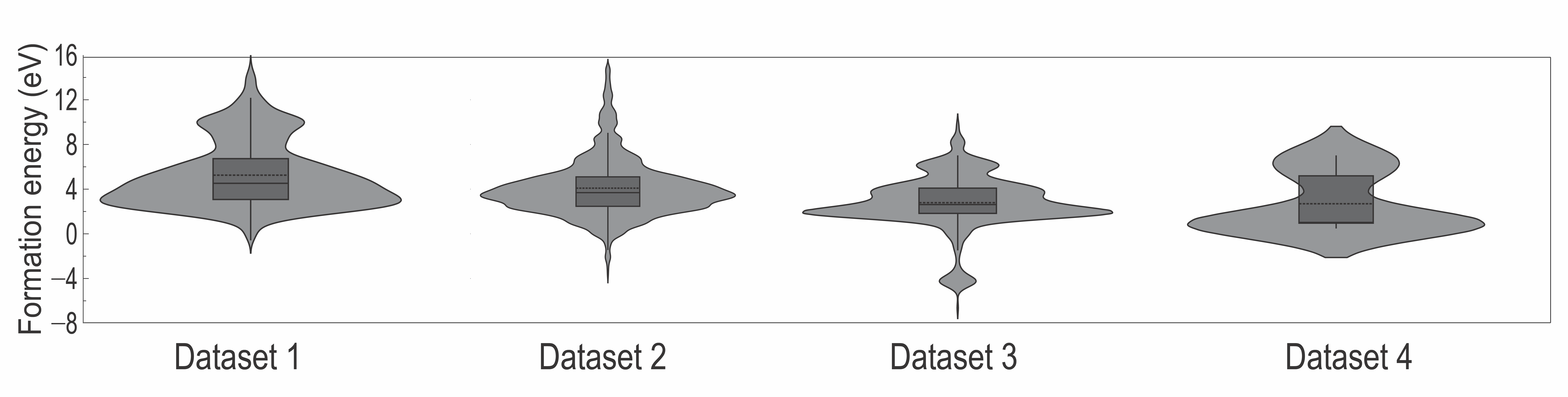}
\caption{\label{fig:violin} Defect formation energy distribution across the four datasets, for neutral defects under A-rich chemical potential conditions.}
\end{figure*}

\section{Computational dataset}

The semiconductor+defect chemical space considered in this work is pictured in \textbf{Fig. S1} in terms of the group IV, III-V, and II-VI binary ZB compounds (referred to henceforth as AB, with A being the cation and B the anion) that serve as hosts to defects, elements selected from across the periodic table as possible defects (henceforth referred to as M for any substitutional or interstitial defect, whereas vacancies will use V), and 5 possible symmetry inequivalent defect sites, namely A-site (M$_{A}$), B-site (M$_{B}$), tetrahedral interstitial site with 4 neighboring A atoms (M$_{i,A}$), tetrahedral interstitial site with 4 neighboring B atoms (M$_{i,B}$), and octahedral interstitial site with 3 A and 3 B atoms in the neighborhood (M$_{i,oct}$). Here, we utilize 4 types of datasets: dataset 1 (all possible single native defects in 34 binary ZB semiconductors), dataset 2 (hundreds of substitutional and interstitial impurities across all ZB compounds), dataset 3 (native defects and impurities in a variety of CdSe$_{x}$Te$_{1-x}$ alloys), and dataset 4 (some defect complexes simulated in CdTe). Datasets 3 and 4 arise from a parallel study on dopant-activation in CdTe-based solar cells \cite{CdSeTe_def} and are used here to evaluate the effectiveness of GNN-based defect predictions for alloys and complexes. All datasets contain DFEs calculated for at least 5 distinct charged states (q = 0, +2, +1, -2, -1) at two extreme chemical potential conditions, namely A-rich and B-rich. \textbf{Fig. \ref{fig:violin}} shows violin plots capturing the distribution of DFEs (only for neutral defects at A-rich chemical potential conditions) for all 4 datasets, with inset box plots showing the median, lower quartile, and upper quartile for each. \\

The DFT details, including specific VASP input parameters, level of theory information, reciprocal space sampling, references for chemical potentials, and equations used for DFE calculation, are present in our past publication \cite{Mannodi-Kanakkithodi2022-ck}. All data is from the semi-local GGA-PBE functional, which generally reproduces lattice parameters and relative stabilities quite well, but is not reliable for electronic band edges and band gaps, which is likely to affect computed defect levels as well. The non-local hybrid HSE06 functional \cite{hse} is preferred for electronic and defect properties, but is much more expensive and often requires tuning of the mixing parameter (which determines the fraction in which exact exchange from Hartree-Fock is combined with the exchange-correlation energy from PBE), which is very system-specific and not trivially applied across large datasets \cite{HP_hse}. Beyond-DFT methods such as the GW approximation, which expands the self-energy in terms of the single-particle Green's function G and the screened Coulomb interaction W \cite{gw}, are more reliable but too prohibitively expensive to be applied high-throughput. In past work \cite{Mannodi-Kanakkithodi2022-ck}, we showed that PBE computed defect charge transition levels, when plotted to span the experimentally-known band gap of the semiconductor, match rather well with measured defect levels for a dataset of $\sim$ 80 defects in binary ZB compounds collected from the literature, showing a PBE vs experiment RMSE of 0.21 eV. \\

Thus, PBE-level DFEs and transition levels may be sufficient for a first-level screening of low-energy defects. Inaccuracies will still persist from incorrectly locating VBM and CBM, but appropriate corrections can be applied afterwards using different higher-accuracy bulk calculations once PBE-level DFEs are predicted for multiple \textit{q} and $\mu$ conditions. Two such possible correction methods include using the modified band alignment approach based on PBE and HSE06 band gap values \cite{mba}, and shifting both band edge positions using GW quasiparticle energies \cite{gw_def}. The focus of present work is to demonstrate the accelerated prediction of PBE-level defect energetics, and the aforementioned corrections will be considered in future work. In the next few subsections, a detailed description is provided for the four datasets generated at the PBE level. \\

\begin{table*}[t]
    \centering
    \begin{adjustbox}{width=1\textwidth}
    \begin{tabular}{|c|c|}
    \hline
        \textbf{Dataset} & \textbf{Data Points} \\ \hline
        Native defects in 34 compounds (Dataset 1) & 2053 (q=+2), 1840 (q=+1), 3071 (q=0), 1966 (q=-1), 1498 (q=-2)  \\ \hline
        Impurities in 34 compounds (Dataset 2) & 5326 (q=+2), 3990 (q=+1), 13966 (q=0), 3568 (q=-1), 4628 (q=-2) \\ \hline
        Native defects in CdSe\textsubscript{x}Te\textsubscript{1-x} (Dataset 3) & 291 (q=+2), 322 (q=+1), 734 (q=0), 305 (q=-1), 329 (q=-2) \\ \hline
        Defect complexes in CdTe (Dataset 4) & 47 (q=0)  \\ \hline
    \end{tabular}
    \end{adjustbox}
    \caption{\label{table:dataset} Number of data points for each charge state q across Datasets 1, 2, 3, and 4.}
\end{table*}

\subsection{Dataset 1}

Dataset 1 contains DFEs for all possible native defects in each AB compound, namely V$_{A}$ (vacancy at A-site), V$_{B}$, A$_{i,A}$ (A self-interstitial at A-coordinated tetrahedral site), A$_{i,B}$, A$_{i,oct}$ (A self-interstitial at octahedral site), B$_{i,A}$, B$_{i,B}$, B$_{i,oct}$, A$_{B}$ (A on B anti-site substitution), and B$_{A}$. All AB compounds, simulated in the cubic ZB structure with A atoms occupying an FCC lattice and B atoms occupying tetrahedral sites, are listed as follows: 8 II-VI compounds (ZnO, ZnS, ZnSe, ZnTe, CdO, CdS, CdSe, and CdTe), 16 III–V compounds (AlN, AlP, AlAs, AlSb, BN, BP, BAs, BSb, GaN, GaP, GaAs, GaSb, InN, InP, InAs, and InSb), and 10 group IV compounds (SiC, GeC, SnC, SiGe, SiSn, GeSn, C, Si, Ge, and Sn). There are a total of 312 possible native defects across the 34 compounds, and DFEs were computed for all under both A-rich and B-rich conditions. From each of the 312 ($\times$ 5 \textit{q} states) PBE geometry optimization calculations, we collected all possible ``intermediate structures'', that is, geometries generated during the course of the optimization, all the way from pristine structure (which is simply the ground state semiconductor bulk supercell structure with a defect introduced) to the fully optimized structure; also collected were the total DFT energies corresponding to each structure. The shell script used to collect intermediate structures (\textit{IS}) for every defect from XDATCAR and corresponding energies from OUTCAR (typical output files in VASP \cite{vasp}) is added to the SI and available on GitHub.  The DFE corresponding to every \textit{IS} is estimated as E$^{f}$(\textit{IS}) = E$_{DFT}$(\textit{IS})- E$_{DFT}$(optimized structure) + E$^{f}$(optimized structure). \\

This approach helps swell the DFT native defect dataset to 3071 data points for q = 0, and between $\sim$ 1500 and $\sim$ 2000 points for the other q values, as reported in \textbf{Table \ref{table:dataset}}. The lower number of structures for the charged systems as compared to the neutral defects comes from the fact that most of the geometry optimization takes place during the neutral calculation, whereas the charged calculations typically use the optimized neutral defect geometry as their starting point. All the \textit{IS} serve as energetically accessible but not ground state defective structures, which can play an important role in understanding the dynamics and behavior of the crystal, but also provide an energetically and structurally diverse dataset for a GNN model to be trained on. This also ensures that the GNN ``knows'' what an unoptimized, partially optimized, and fully optimized defect structure looks like, meaning it will yield the correct energy corresponding to any hypothetical new structure and potentially help reduce the energy by subtly modifying the defect coordination environment. \\

\subsection{Dataset 2}

Dataset 2 contains hundreds of impurities or extrinsic defects (M) at various sites, namely M$_{A}$, M$_{B}$, M$_{i,A}$, M$_{i,B}$, and M$_{i,oct}$, across each of the 34 unique AB compounds. The five distinct defect sites are considered in 30 binary compounds and three defect sites (A-site, one tetrahedral interstitial site, and one octahedral interstitial site) are considered in the elemental systems (C, Si, Ge, and Sn). This dataset encompasses a wide range of singular impurity atoms, including groups I to VII, all transition metals, and all lanthanides, leading to a total of 77 species, as shown in \textbf{Fig. S1}. The total number of possible impurities resulting from this can be calculated as 77 × 5 × 30 + 77 × 3 × 4 = 12,474 (many of these would coincide with the native defects described earlier). Out of this dataset of 12,474 defects, 1566 were chosen for complete neutral-state geometry optimization, and $\sim$ 1000 were subjected to charged calculations as well; points in the DFT dataset exhibit sufficient chemical diversity in terms of semiconductor type, element type, and defect site type, to adequately represent the entire chemical space. Once again, we collected all IS from the DFT optimization runs for each impurity in 5 different q states, leading to nearly 14,000 data points for q=0 and between 3500 and 5300 data points for the other q values, as reported in \textbf{Table \ref{table:dataset}}.

\begin{figure*}[htp]
\centering
\includegraphics[width=.9\linewidth]{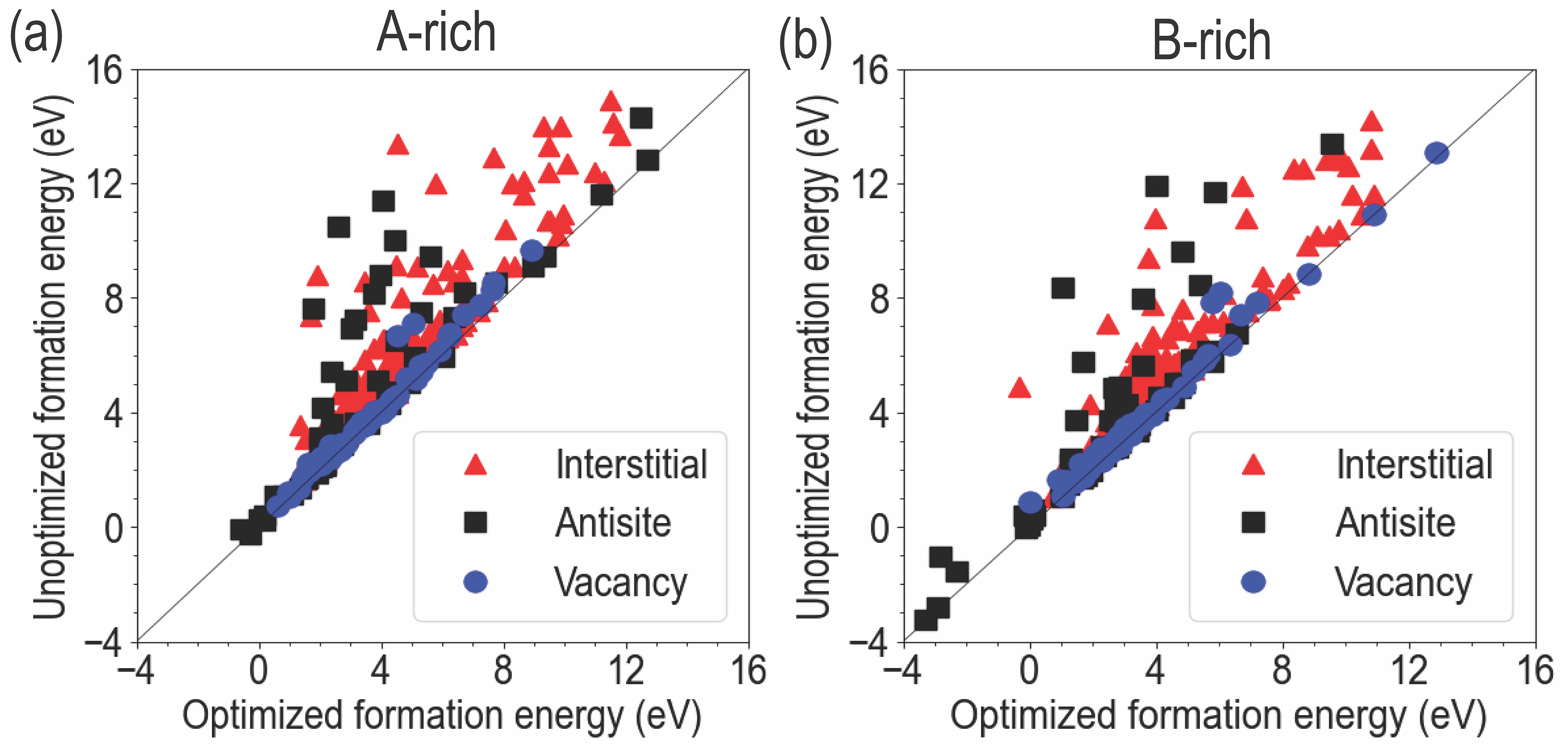}
\caption{\label{fig:opt-unopt} DFT unoptimized vs optimized neutral defect formation energy in Dataset 1, under (a) A-rich, and (b) B-rich chemical potential conditions.}
\end{figure*}

\subsection{Dataset 3}

This dataset includes several possible native defects in a series of CdSe$_{x}$Te$_{1-x}$ alloys (x = 0.125, 0.25, 0.375, 0.5, 0.625, 0.75, and 0.875), namely V$_{Cd}$, V$_{Te}$, V$_{Se}$, Cd$_{i}$, Te$_{i}$, Se$_{i}$, Cd$_{Te}$, Cd$_{Se}$, Te$_{Cd}$, and Se$_{Cd}$. This results in a total of 82 unique defects across the 7 mixed compositions, which are of interest in CdTe solar cells where Se is often mixed at the anion site \cite{Li2017-ez,Poplawsky2016-hq,CdTe_1,CdTe_2,CdTe_3}. DFEs are computed for each defect in 5 q states at Cd-rich and Te/Se-rich conditions to obtain the two extreme ranges of energies, and all the \textit{IS} are collected from the optimization runs. Total datasets of a few hundred points are thus compiled, as shown in \textbf{Table \ref{table:dataset}}. This dataset will help examine whether GNN models trained on defects in pure unmixed compositions (CdTe and CdSe) are applicable to alloyed compositions in the same chemical space, and how many new alloy data points might need to be added to the training set to achieve satisfactory accuracy. \\

\subsection{Dataset 4}

Finally, we posit that crystal graphs should be capable of representing any type of defect complexes as well in addition to the single-point defects described above. For exhaustively investigating defect tolerance and dopability of a semiconductor, it is vital to consider likely defect complexes, which are typically multiple point defects or impurities that form simultaneously in the lattice and interact with each other. Examples include Schhotky and Frenkel defects, and compensating vacancies or interstitials that form along with dopants. The V$_{Ga}$–O$_{N}$–2H triple complex was found to have a very low energy in GaN \cite{Wickramaratne}, and it has recently been suggested that As-O based complexes may form in CdTe \cite{CdSeTe_def}. Thus, we simulated a series of complexes such as V$_{Cd}$+As$_{Te}$ and V$_{Te}$+Cu$_{Cd}$ in CdTe, resulting in a small dataset of 47 points of neutral state defects, for both Cd-rich and Te/Se-rich conditions, including all the \textit{IS}.

\section{DFT optimized vs unoptimized formation energy }

Before training GNN models, we analyzed the DFT datasets to determine the scale of the differences between DFEs from full DFT-optimization and from pristine, unoptimized defect structures. For any hypothetical defect, an unoptimized pristine structure could be trivially generated simply by inserting the defect of interest in an optimized bulk supercell, but obtaining the ground state DFE is a matter of optimizing this structure, which would involve a PBE calculation that runs for minutes, hours, or days, depending on the nature of the defect. The purpose of GNN-based on-demand predictions is to reduce this time drastically. Since any new GNN predictions would likely be made using pristine defect structures, it is informative to examine, for a given type of defect, how low the energy could go starting from the unoptimized DFE if full optimization were to be performed. \\

\textbf{Fig. \ref{fig:opt-unopt}} shows unoptimized DFE plotted against the fully optimized DFE, for the dataset of 312 native defects across 34 AB compounds (Dataset 1), at both A-rich and B-rich conditions. The unoptimized DFEs are obtained by performing fixed-volume single ionic step calculations on the pristine defect-introduced structures. The dataset is divided into vacancies, anti-site substitutions, and self-interstitials. It can be seen that the amount of geometry relaxation happening in vacancy structures is minimal, with the two types of DFEs almost always being very similar to each other. On the other hand, interstitials and anti-sites often undergo major atomic rearrangement, such that the optimized defect coordination environment may look starkly different from the starting structure, thus leading to DFE differences ranging from 1 eV to nearly 8 eV. The large differences for interstitials could be understood in terms of the unfavorability of introducing an extra cation or anion in a tetrahedral or octahedral void; the larger the ionic radii, the greater this unfavorability. Substitutions depend on the size differences between A and B, and may thus show either a low or high energy variability. These trends roughly inform the threshold that must be applied upon unoptimized DFEs (say, from GNN predictions) to determine their likelihood of stability upon full optimization. It should be noted that the intermediate structures collected from each ``optimization run'' essentially span the range of the unoptimized to optimized DFE, yielding hundreds of structures for some defects and only a handful for others. \\

\section{Graph Neural Network Architecture}

In this section, we briefly describe the technical details behind the three GNN schemes chosen in this study, namely CGCNN, MEGNET, and ALIGNN. \\

\subsection{Crystal Graph Convolutional Neural Network (CGCNN)}

CGCNN, developed by Xie et al. \cite{CGCNN}, is a deep learning architecture that takes a crystal graph as input and applies a series of convolution and pooling operations to extract features that capture the underlying properties of the crystal. These features are subsequently fed into a fully connected neural network (FCNN) to make predictions of the properties of interest\cite{Lee_Asahi}. The CGCNN framework is pictured in \textbf{Fig. S2(a)} and its operation is described below:

\begin{enumerate}

    \item Structure representation: The crystal structure is represented as a graph where the atoms are nodes and the bonds are edges. The nodes and edges are labeled with features such as atomic coordinates, and bond length. 

    \item Graph convolutional layers: CGCNN applies multiple graph convolutional layers to the input graph wherein each layer aggregates information from neighboring nodes and edges and learns features that capture the local environment. Generally, the convolution operation includes computing a weighted sum of the features of neighboring nodes and edges, followed by a non-linear activation function\cite{Kipf_Welling_2016, Zhou_Cui} :   
    \begin{equation}
    H^{(l+1)}=\sigma\left(\sum_{j \in \mathcal{N}(i)} W^{(l)} h_j^{(l)}+W^{(l)} h_i^{(l)}\right)
    \end{equation}    
    Here, $H^{(l+1)}$ is the output feature matrix of the $(l+1)$-th layer, $W^{(l)}$ is the weight matrix of the $l$-th layer, $\sigma$ is a non-linear activation function, $h_i^{(l)}$ is the feature vector of node $i$ in layer $l$, and $\mathcal{N}(i)$ is the set of neighboring nodes of node $i$ in the graph.

    \item Pooling layer: The output of the last convolutional layer is passed through a global pooling layer (e.g., min, max pooling), which aggregates the features of all nodes in the graph into a single vector\cite{Kipf_Welling_2016, Zhou_Cui}. This vector contains information about the entire crystal structure, including all atomic coordinates, bond distances, and well-known elemental properties of each atom such as ionization energy and electronegativity.
    \begin{equation}
    h_{\text {pool }}=\frac{1}{N} \sum_{i=1}^N h_i^{(L)}
    \end{equation}    
    Here, $h_{pool}$ is the output feature vector of the pooling layer, $N$ is the total number of nodes in the graph, and $h_i^{(L)}$ is the feature vector of node $i$ in the last layer $L$.

    \item Fully connected neural network (FCNN): Finally, the output of the pooling layer is fed into an FCNN, which is trained like a regular NN-regression model to make predictions.    
    \begin{equation}
    y=f\left(W_{f c} h_{p o o l}+b_{f c}\right)
    \end{equation}    
    Here, $y$ is the predicted property, $W_{fc}$ is the weight matrix of the FCNN, $b_{fc}$ is the bias vector, $h_{pool}$ is the output feature vector of the pooling layer, and $f$ is a non-linear activation function such as ReLU or sigmoid.
 
\end{enumerate}

\subsection{Materials Graph Network (MEGNET)}

The MEGNET framework was developed and released by Chen et al. in 2019 \cite{MEGNET} and is pictured in \textbf{Fig. S2(b)}. MEGNET uses elemental embeddings to encode periodic chemical trends that can be used to improve the performance of models with limited training data. Elemental embeddings are vector representations of elements that capture their chemical properties that are typically learned from a large dataset of crystals. When a new crystal is encountered, the embeddings can be used to predict the properties of interest. The MEGNET methodology is described below:

\begin{enumerate}

\item Graph representation of materials: MEGNET represents the crystal as a graph where the atoms are the nodes, and the edges represent the connections between the atoms. Each atom is associated with a set of features such as its atomic number, coordinates, and chemical environment.

\item Message passing algorithm: MEGNET uses a message-passing algorithm to capture the interactions between atoms in the crystal. Each atom sends a message to its neighboring atoms, which is a function of the node and edge features. The messages are then aggregated at each node and the resulting feature vector is used as input to a neural network. 

\item Readout layer: The output of the message-passing algorithm is passed through a readout layer which maps the learned node and edge features to target properties, and a loss function is calculated to capture the difference between the predicted and actual values. 

\end{enumerate}

\subsection{Atomistic Line Graph Neural Network (ALIGNN) }

ALIGNN is a novel approach developed by Choudhary et al. \cite{ALIGNN}, that differs from CGCNN and MEGNET in terms of considering three-body interactions (bond angles) as well in addition to two-body terms (bond lengths). ALIGNN leverages both graph convolutional layers and line graph convolutional \cite{Chen_Li_Bruna_2017} layers to capture both short-range and long-range correlations in the crystal. The framework (pictured in \textbf{Fig. S2(c)}) is described below: 

\begin{enumerate}

\item Atomic feature extraction: ALIGNN takes as input a graph representing the atomic structure of the crystal. Each node (atom) in the graph is associated with a set of atomic features, which includes properties such as electronegativity, group number, covalent radius, valence electrons, ionization energy, electron affinity, atomic block, and atomic volume. Each edge (bond) in the graph is associated with both the bond angle and bond distance.

\item Graph convolutional layers: ALIGNN uses graph convolutional layers to update the feature vectors of each node based on the features of its neighboring nodes. In each layer, the feature vectors are updated using a weighted sum of the features of neighboring nodes, similar to other models. This step captures short-range interactions in the structure.

\item Line graph construction: To capture long-range correlations, ALIGNN constructs a line graph on top of the original crystal graph. The line graph has nodes that represent unique bonds between atoms, corresponding to edges in the crystal graph. The line graph edges connect pairs of bonds that share a central atom in the crystal graph, effectively capturing bond angle information. ALIGNN then applies another set of graph convolutional layers to the line graph, which updates the feature vectors of each bond based on the features of neighboring bonds. The information from the line graph is then propagated back to the original crystal graph to update the bond features in combination with the node features.

\item Feature refinement: After the line graph convolution, ALIGNN refines the feature vectors for each edge using a set of learnable transformations that help capture more complex interactions between atoms and bonds.

\item Graph pooling: ALIGNN aggregates the refined bond-level feature vectors into a single graph-level feature vector using a graph pooling operation that summarizes the relevant information from the entire crystal graph.

\item Output prediction: Finally, the graph-level feature vector is fed to an FCNN for making predictions.

\end{enumerate}

\begin{figure*}[t]
\centering
\includegraphics[width=1.0\linewidth]{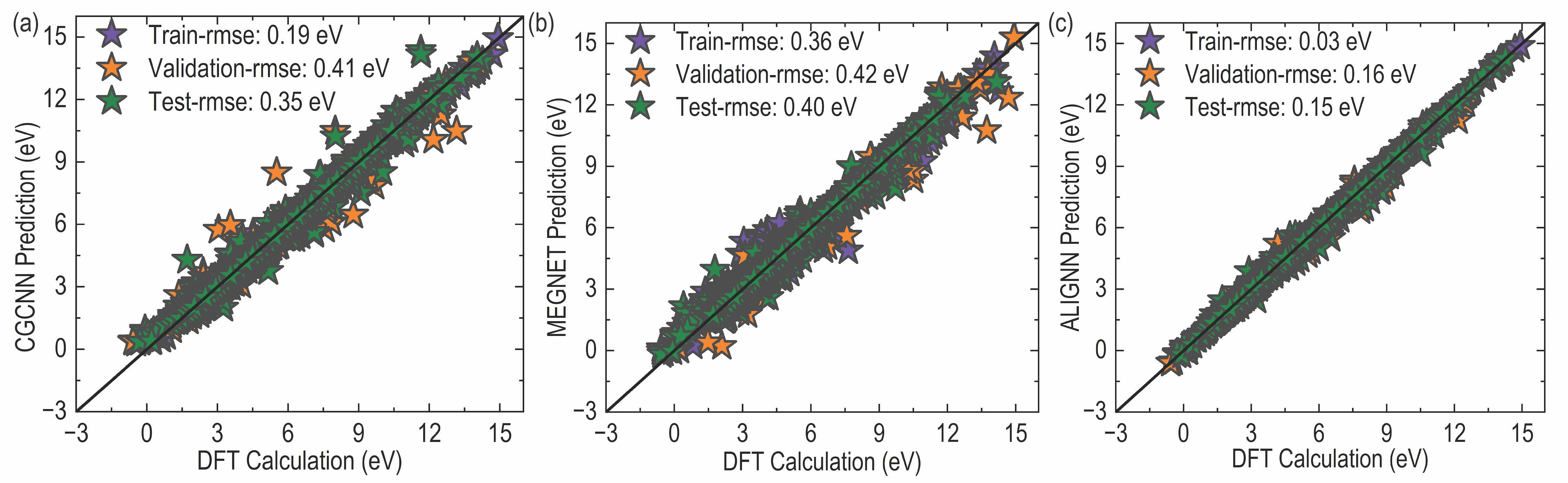}
\caption{\label{fig:gnn_A} Parity plots for rigorously optimized (a) CGCNN, (b) MEGNET, and (c) ALIGNN models, trained on Dataset 1 for A-rich chemical potential conditions and q = 0. }
\end{figure*}

\section{Results and Discussion}

\subsection{Testing GNN models on Dataset 1}

As a first step, we tested the performance of CGCNN, MEGNET, and ALIGNN for predicting the q=0 E$^{f}$ for Dataset 1 only. For each model, the data is split 60:20:20 into training, validation, and test sets. The CGCNN training protocol has several important hyperparameters that must either be kept fixed at recommended values or rigorously tuned over a set of possible values, such as the number of properties used in the atom feature vector, the number of convolutional layers, the length of the learned atom feature vector, the number of hidden layers, the regularization term, the scaling factor of the Gaussian initialization of weights, step size of the Adam optimizer~\cite{kingma2014adam}, dropout fraction, batch size, number of epochs, and the cut-off distance r$_{c}$ for constructing the crystal graph. Here, we optimized only the batch size, epochs and r$_{c}$, keeping the rest of the hyperparameters the same as in the original CGCNN publication \cite{CGCNN}. A parity plot for the best CGCNN predictions on Dataset 1 (for A-rich conditions) is pictured in \textbf{Fig. \ref{fig:gnn_A}(a)}, showing RMSE of 0.19 eV, 0.41 eV, and 0.35 eV respectively on the training, validation, and test sets. These errors are already far lower than our previously published DFE test prediction errors of $\sim$ 0.6 eV for defects in Cd-based chalcogenides \cite{Mannodi_Toriyama_2020} and $\sim$ 1 eV across many group IV, II-VI, and III-V semiconductors \cite{Mannodi-Kanakkithodi2022-ck}, as well as being highly competitive with the state of the art for such predictions. Learning curves showing how the CGCNN errors converge as the epochs, batch size, and r$_{c}$ increase are presented in \textbf{Fig. S3}. \\

Next, we trained a MEGNET model as shown in \textbf{Fig. \ref{fig:gnn_A}(b)} following the same strategy. Notable hyperparameters include the number of interaction blocks, number of hidden layers, hidden layer size, learning rate, regularization strength, dropout rate, batch size, activation function, number of features assigned to each bond in the input graph, and r\textsubscript{c}. Here, we only optimized the number of epochs, batch size, and r\textsubscript{c}, and the rest of the parameters are directly adopted from the MEGNET publication \cite{MEGNET}. We find that MEGNET once again performs much better than previous composition-based models, but shows slightly larger errors than CGCNN with RMSE of 0.36 eV, 0.42 eV, and 0.40 eV on the training, validation, and test sets, respectively. The test error is close enough to the CGCNN error of 0.35 eV to justify the use of MEGNET over CGCNN, especially since MEGNET significantly corrects any possible overfitting in the CGCNN models by yielding roughly similar training, validation, and test errors. MEGNET has a more complex model architecture than CGCNN and includes elemental embeddings encoding periodic chemical trends which may allow better generalization to unseen data. 

Finally, we trained an ALIGNN model on Dataset 1 and found that it yields better performance than both CGCNN and MEGNET, with a slight concern of model overfitting alleviated by the remarkably low values of validation and test RMSEs. As shown in \textbf{Fig. \ref{fig:gnn_A}(c)}, the test RMSE is 0.15 eV, which represents a 99~\% accuracy considering the DFE values range from 0 eV to 15 eV. The reason for the improvement from CGCNN and MEGNET could be attributed to the line graph convolution step that captures long-range interactions, which may be important for particular point defects whose presence affects atomic arrangements beyond just the first nearest neighbor shell, causing larger lattice distortions. For training ALIGNN models, we use r$_{c}$ = 8 $\text{\AA}$, 12 nearest neighbors, 4 graph convolutional layers, 4 line graph layers, learning rate = 0.001, batch size = 16, an Adamw optimizer, and 150 epochs. Results of hyperparameter optimization with ALIGNN are presented in \textbf{Fig. S4}, which is a more computationally expensive step than for the other two methods, motivating the use of much of the same hyperparameters as in past work \cite{ALIGNN}. However, the training time is still reasonable and the accuracy payoff is immense; thus, we use ALIGNN as the GNN scheme of choice going forward, and discuss prediction performances for Datasets 2, 3, and 4 in the next subsection. \textbf{Table \ref{table:RMSE}} lists the optimized training, validation, and test set RMSE values for CGCNN, MEGNET, and ALIGNN models trained on Dataset 1. \\

\begin{table*}[t]
    \centering
    \begin{adjustbox}{width=0.6\textwidth}
    \begin{tabular}{|c|c|c|c|}
    \hline
        \textbf{GNN Scheme} & \textbf{Train RMSE (eV)} & \textbf{Validation RMSE (eV)} & \textbf{Test RMSE (eV)} \\ \hline
        CGCNN & 0.19 & 0.41 & 0.35 \\ \hline
        MEGNET & 0.36 & 0.42 & 0.40 \\ \hline
        ALIGNN & 0.03 & 0.16 & \textbf{0.15} \\ \hline
    \end{tabular}
    \end{adjustbox}
    \caption{\label{table:RMSE} Training, validation, and test set RMSEs for different GNN models trained on Dataset 1 at A rich conditions.}
\end{table*}

\begin{figure*}[t]
\centering
\includegraphics[width=1.0\linewidth]{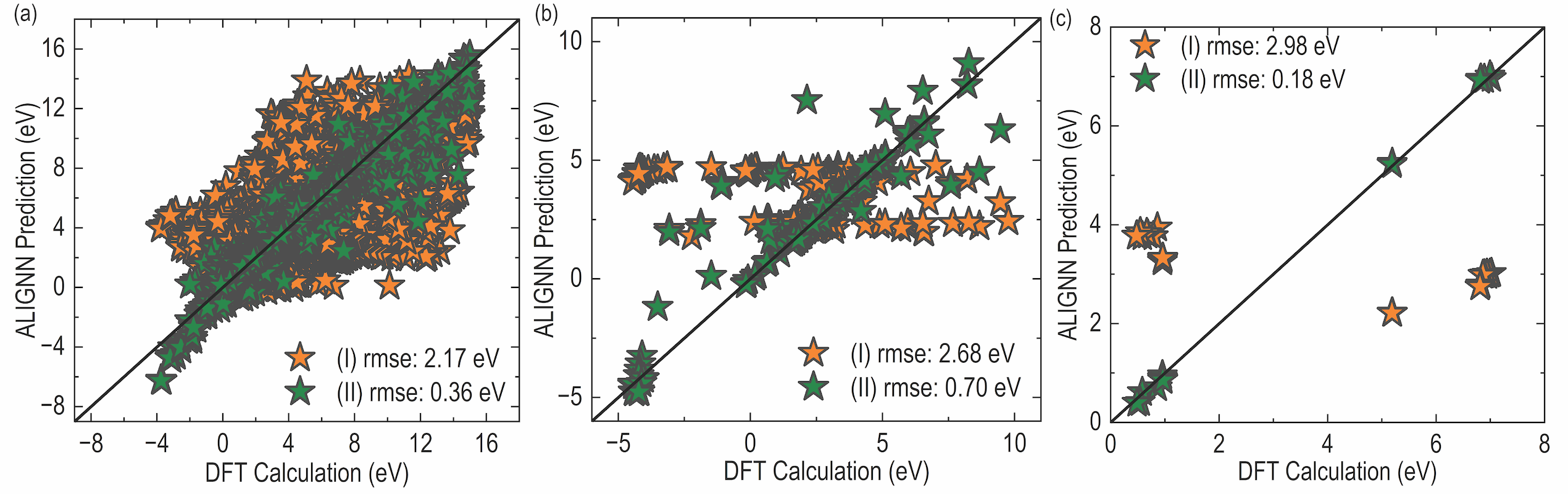}
\caption{\label{fig:gnn_B} Parity plots for ALIGNN models trained on (q=0, A-rich) (a) Dataset 1 + Dataset 2, (b) Dataset 1 + Dataset 3, and (c) Dataset 1 + Dataset 4. (I) refers to models trained purely on Dataset 1 and tested on Dataset 2, 3 or 4, whereas (II) refers to 50~\% of the new dataset added to the training set with Dataset 1.}
\end{figure*}

\subsection{Extending ALIGNN to Datasets 2, 3, and 4}

To determine whether the GNN models optimized so far could directly be applied to impurities, complexes, and alloys, we first tested the ALIGNN model trained on dataset 1 for their predictive power on datasets 2, 3, and 4, before suitably re-optimizing the models by adding more training data points. \textbf{Fig. \ref{fig:gnn_B}} shows the prediction performance of the ALIGNN model trained only on dataset 1 for dataset 2 (a), dataset 3 (b), and dataset 4 (c), along with the improvement in the predictions when 50~\% of any new dataset is added to the training set and the ALIGNN model is retrained. The RMSE for the dataset 1-trained ALIGNN model is as high as 2.17 eV for dataset 2, 2.68 eV for dataset 3, and 2.98 eV for dataset 4, showing very poor performances that become worse going from native defects to impurities to alloys to complexes. The structure-property relationships learned from native defects alone cannot be generalized to extrinsic species or non-binary compounds, or indeed, the presence of multiple simultaneous defects that will inevitably cause far higher distortions and atomic rearrangements compared to single defects. \\

Upon combining 50~\% of each dataset (chosen randomly) with dataset 1 and re-optimizing the model using the same approach as before, and performing necessary hyperparameter optimization anew, RMSE values improve drastically to 0.36 eV for dataset 2, 0.70 eV for dataset 3, and 0.18 eV for dataset 4. These errors will go further down as more data is added to the training set, showing that each type of defect data needs to be adequately represented during the training process for generalizing the ALIGNN predictive power. This exercise provides insights into the challenges associated with representing and predicting the energetics of defects in different defect categories with a limited training dataset and demonstrates the importance of training ML models on comprehensive datasets to improve their performance across various defect types.  \\

\begin{figure*}[t]
\centering
\includegraphics[width=1.0\linewidth]{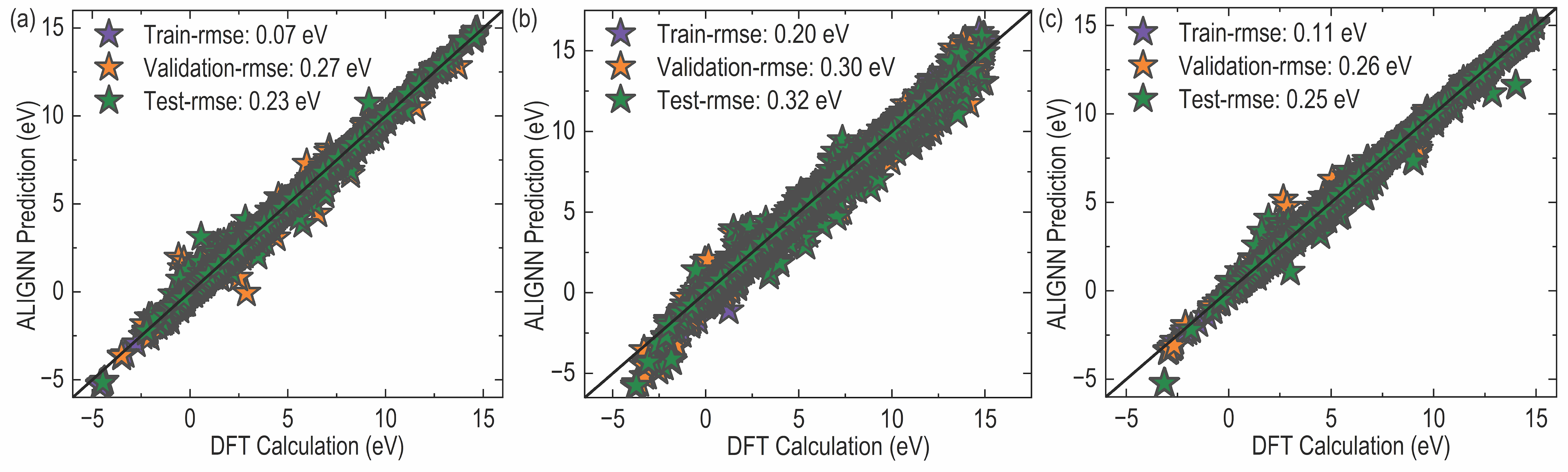}
\caption{\label{fig:gnn_comb} Parity plots for ALIGNN models trained on Datasets 1 + 2 + 3 + 4, for (a) E$^{f}$(q=+1, E$_{F}$=0), (b) E$^{f}$(q=0, E$_{F}$=0), and (c) E$^{f}$(q=-1, E$_{F}$=0), under A-rich chemical potential conditions. Parity plots for E$^{f}$(q=+2, E$_{F}$=0) and E$^{f}$(q=-2, E$_{F}$=0), as well as all parity plots for B-rich DFEs, are presented in the SI.}
\end{figure*}

Next, we trained predictive models by combining all four datasets, for all charge states and chemical potential conditions. For charged defects, the DFE value is taken to be E$^{f}$(E$_{F}$=0), because by predicting this value for each charge state, the E$^{f}$ vs E$_{F}$ plot can be extended across the band gap region by using straight lines with slope = q. Thus, a total of 10 different ALIGNN models are trained for the combined dataset, for E$^{f}$(q=+2, E$_{F}$=0), E$^{f}$(q=+1, E$_{F}$=0), E$^{f}$(q=0, E$_{F}$=0), E$^{f}$(q=-1, E$_{F}$=0), and E$^{f}$(q=-2, E$_{F}$=0), at A-rich and B-rich chemical potential conditions. As seen in \textbf{Fig. S5}, there are a handful of outliers in the parity plots, especially in the case of E$^{f}$(q=0, E$_{F}$=0), which may result from some structures getting stuck in local minima during DFT optimization, and possible inaccuracies in the GNN model that may be fixed with more data and/or hyperparameter optimization. We removed a few notable outliers and trained the models again, to obtain the best ALIGNN predictive models that may be applied for new data points. The q=+1, q=0, and q=-1 ALIGNN models at A-rich conditions are pictured in \textbf{Fig. \ref{fig:gnn_comb}(a)}, \textbf{(b)}, and \textbf{(c)}, respectively, and the remaining 7 models are presented in \textbf{Figs. S6} and \textbf{S7}. These models show very respectable RMSE values, suggesting that ALIGNN is capable of effectively learning the structure-property correlations in each dataset and each charge state, and generalizing them across all data types. Test prediction errors for q=+2, q=+1, q=0, q=-1, and q=-2 are found to be 0.30 eV, 0.23 eV, 0.32 eV, 0.25 eV, and 0.26 eV, respectively, representing a 98~\% accuracy. The slightly larger errors for the neutral defects arise from the larger structural diversity for q=0 defect structures compared to the charged defect structures, also manifesting in much larger numbers of q=0 data points (e.g., 13,966 in dataset 2) than q=+1 (3990 in dataset 2) or q=-1 (3568 in dataset 2). The training, validation, and test errors for the best ALIGNN models for different charge states under A-rich conditions are listed in \textbf{Table \ref{table:q_rmse}}. \\

\begin{table*}[t]
    \centering
    \begin{adjustbox}{width=0.6\textwidth}
    \begin{tabular}{|c|c|c|c|}
    \hline
        \textbf{Charge} & \textbf{Train RMSE (eV)} & \textbf{Validation RMSE (eV)} & \textbf{Test RMSE (eV)} \\ \hline
        q = +2 & 0.10 & 0.25 & 0.30 \\ \hline
        q = +1 & 0.07 & 0.27 & 0.23 \\ \hline
        q = 0 & 0.20 & 0.30 & 0.32 \\ \hline
        q = -1 & 0.11 & 0.26 & 0.25 \\ \hline
        q = -2 & 0.06 & 0.26 & 0.26 \\ \hline
    \end{tabular}
    \end{adjustbox}
    \caption{\label{table:q_rmse} Training, validation, and test set RMSEs for ALIGNN models trained on different charge states at A rich condition.}
\end{table*}

\begin{figure*}[t]
\centering
\includegraphics[width=.9\linewidth]{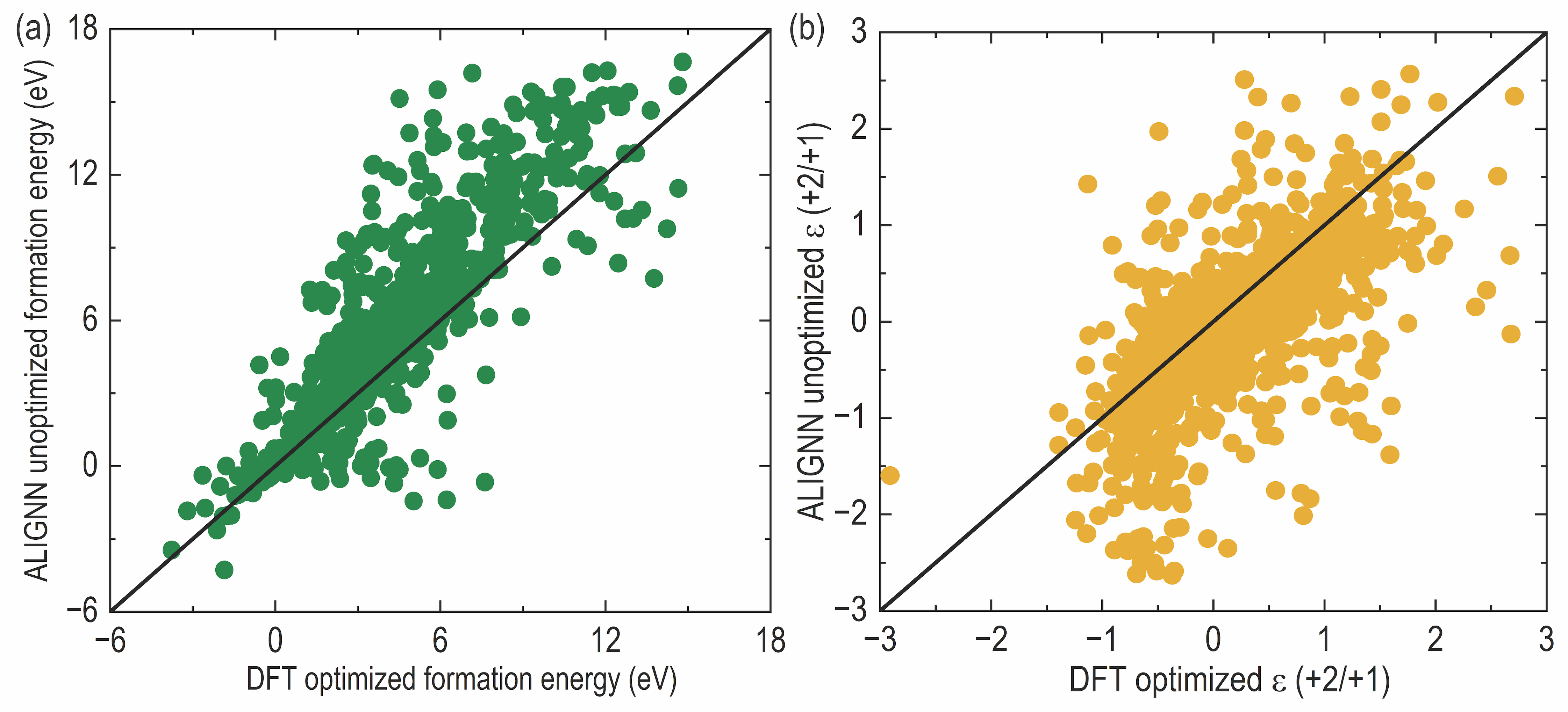}
\caption{\label{fig:gnn_ctl} (a) ALIGNN-unoptimized vs DFT-optimized E$^{f}$(q=0, E$_{F}$=0) under A-rich chemical potential conditions. (b) ALIGNN-unoptimized vs DFT-optimized $\epsilon$(+2/+1) charge transition levels. All data are shown for combined datasets 1 + 2 + 3 + 4.}
\end{figure*}

\begin{figure*}[t]
\centering
\includegraphics[width=.9\linewidth]{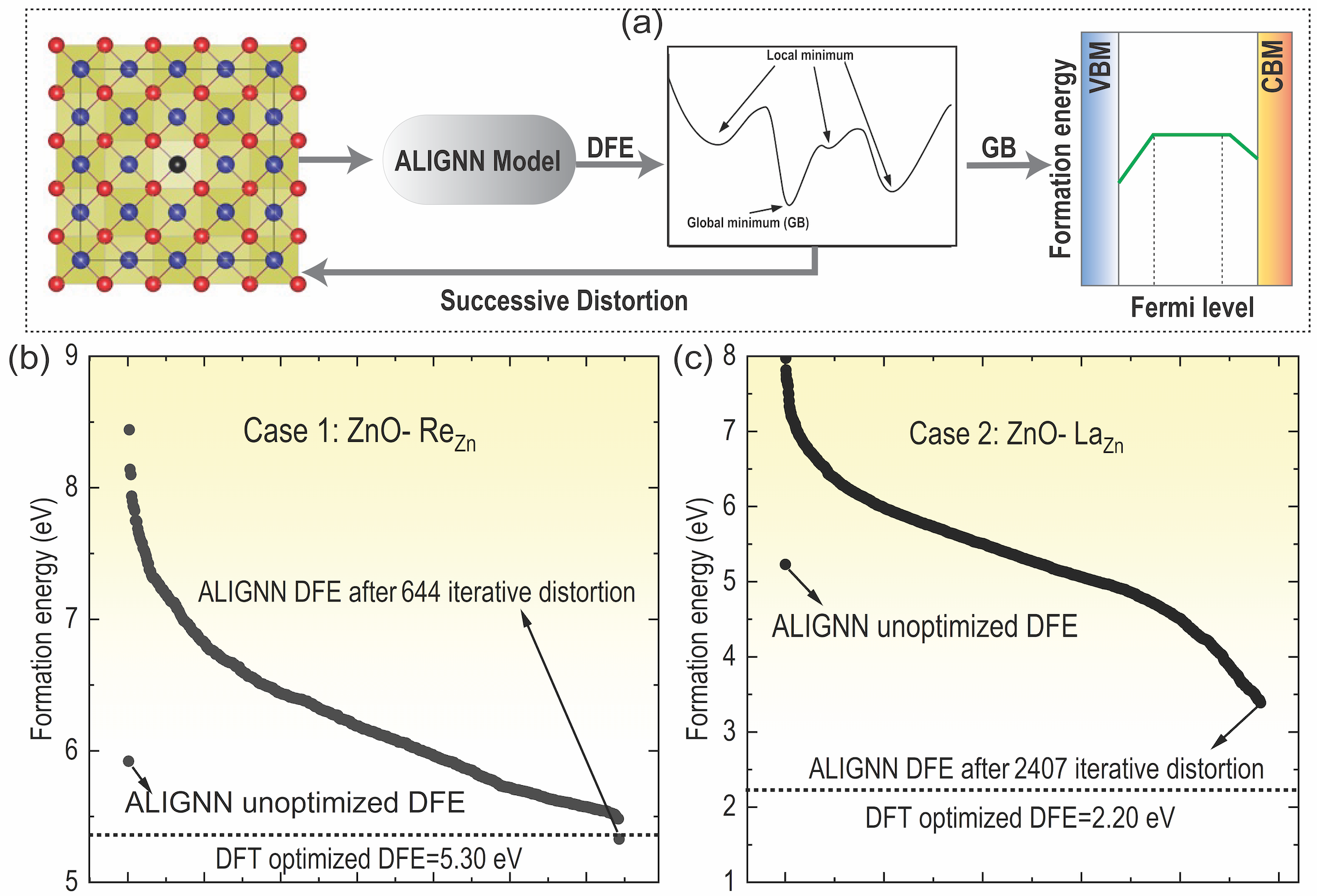}
\caption{\label{fig:alignn_opt} (a) ALIGNN-based defect structure optimization scheme, demonstrated for Re$_{Zn}$ (b) and La$_{Zn}$ (c) in ZnO under Zn-rich chemical potential conditions.}
\end{figure*}

\begin{figure*}[t]
\centering
\includegraphics[width=1\linewidth]{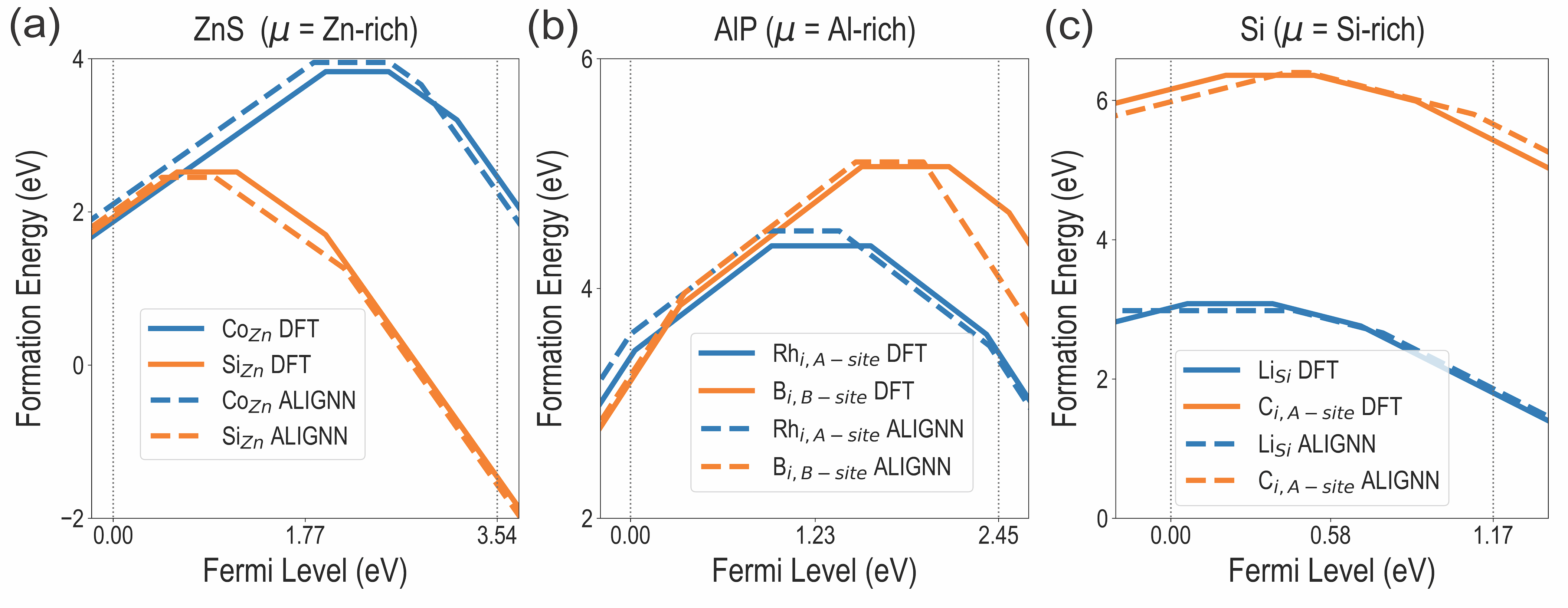}
\caption{\label{fig:forme} ALIGNN-optimized and DFT-optimized defect formation energy plots for two selected defects each in (a) ZnS under Zn-rich conditions, (b) AlP under Al-rich conditions, and (c) Si under Si-rich conditions.}
\end{figure*}

\begin{figure}[t]
\centering
\includegraphics[width=1\linewidth, height=8.5cm]{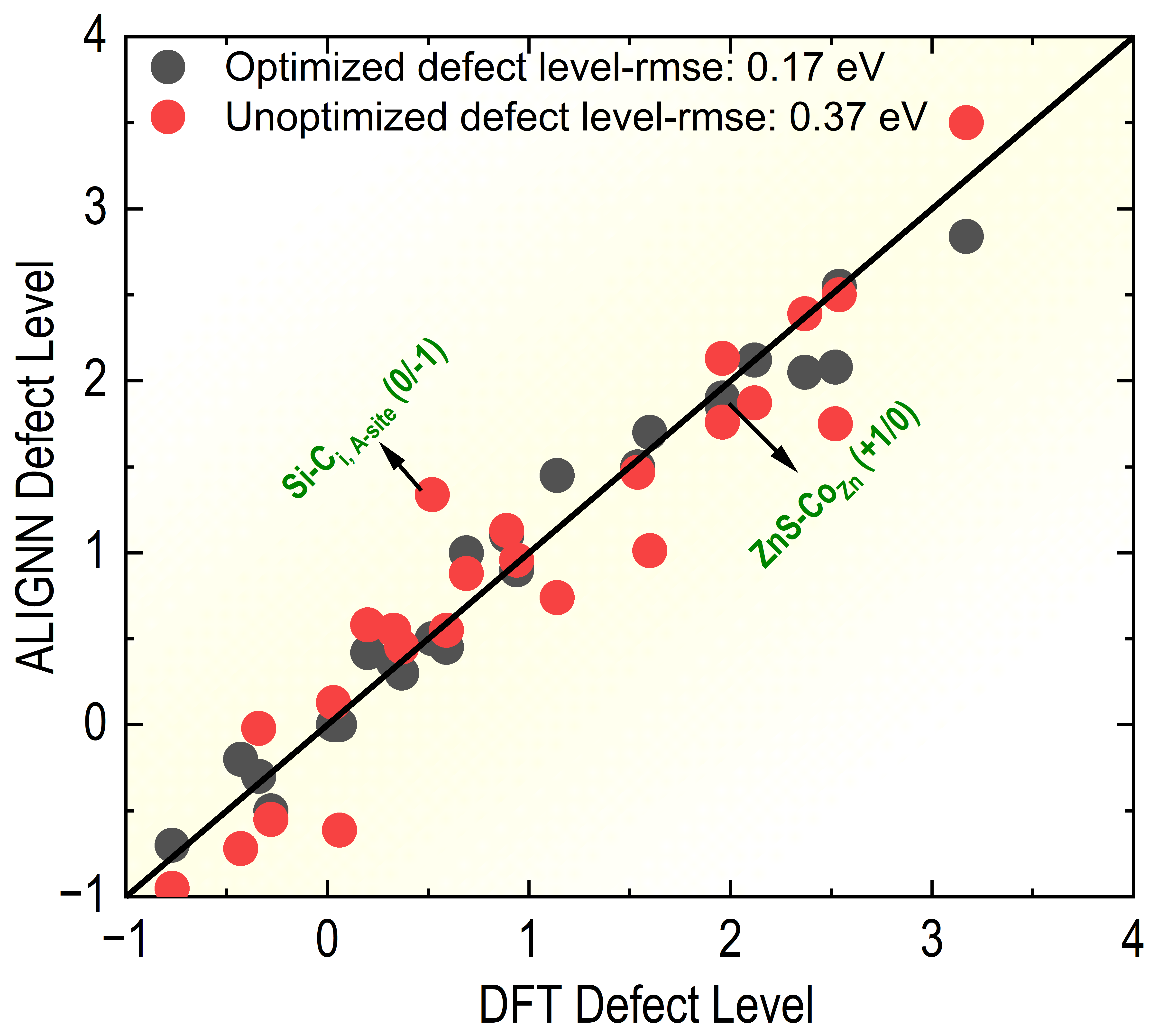}
\caption{\label{fig:op_ctl} ALIGNN-optimized and ALIGNN-unoptimized defect charge transition levels plotted against corresponding DFT-optimized values.}
\end{figure}

\subsection{ALIGNN-unoptimized vs DFT-optimized energies}

The next objective is to utilize the best ALIGNN models to make predictions for new defects and perform screening of potentially low-energy defects based on predicted DFEs. The caveat here is that for any new defect, one could only really generate an ``unoptimized'' defect structure, and thus the only ALIGNN prediction that can be made is the unoptimized DFE. As described earlier, a full DFT optimization of any defect structure is obviously a time-consuming step, involving structural relaxation until atomic forces become close to zero and the energy of the crystal reaches a minimum. In contrast, ALIGNN prediction of unoptimized DFE can be performed in seconds and thus used to estimate the unoptimized energies of hundreds of thousands of defect structures, which could then be used as a surrogate for screening based on some upper bound values \cite{ube1,ube2}. However, ALIGNN predictions could also, in theory, be used to replace DFT-based energy estimates within a gradient descent-type optimization process \cite{Choudhary_DeCost,SchNet,m3gnet}, or using brute-force structure generation and energy evaluation, and thus quickly yield low energy structures at near-DFT accuracy for any hypothetical crystalline defects. \\

To test the correspondence between ALIGNN-unoptimized energies and DFT-optimized energies, we plotted the ALIGNN-predicted E$^{f}$(q=0) (ALIGNN-unopt) on pristine structures of all defects across datasets 1, 2, 3, and 4, against the DFT-optimized E$^{f}$(q=0) (DFT-opt), in \textbf{Fig. \ref{fig:gnn_ctl} (a)}. We expect the ALIGNN-unopt DFE values to always be higher than the DFT-opt values, and this is indeed the case of $>$ 95~\% of the data points. However, we do find the opposite to be true for several defects, which is perhaps a consequence of the statistical nature of ML predictions, which are very accurate on average but may show large errors for certain outliers. Importantly, notable differences between ALIGNN-unopt and DFT-opt values are seen for many defects where large structural changes are expected upon full relaxation. Some examples include V$_{Si}$ in SiC (8.28 eV difference), V$_{In}$ in InN (7.63 eV difference), and Cs$_{i,oct}$ in SiC (7.17 eV difference). By examining the 300 defects out of this total list of 1747 defects (plotted in \textbf{Fig. \ref{fig:gnn_ctl} (a)}) which show the largest ALIGNN-unopt vs DFT-opt energy differences, we find an average difference of $\sim$ 1 eV; thus, we could apply a rough general equivalence of DFT-opt = ALIGNN-unopt -- 1 eV, and use this information to perform a high-throughput screening of likely low energy defects. Similar trends are expected to hold for ALIGNN-unopt vs DFT-opt energies of charged defects as well. \\

Looking at only the DFT-opt values, we find that 170 defects have DFE < 0 eV, with some examples including N$_{S}$ in ZnS (-3.76 eV), N$_{As}$ in GaAs (-3.2 eV), and N$_{Te}$ in CdTe (-2.64 eV). Whereas a look at the ALIGNN-unopt values yields 351 defects with DFE < 1 eV which includes all 170 low energy defects from DFT, meaning that the suggested upper-bound energy screening procedure should help account for all potentially low energy defects in addition to a few unstable defects which may be eliminated subsequently. On average, the computational cost for optimizing a single-point defect in a 64-atom supercell amounts to approximately 2500 core hours for the neutral state and 1000 core hours each for charged calculations. Running only static, single-shot calculations on pristine structures requires around 400 core hours in total. On the other hand, ALIGNN-unopt predictions are made in seconds with minimal computing expense, and it is imperative to use these predictions as descriptors for low energy defects. Visualizing the ALIGNN-unopt vs DFT-opt values for q = +2, +1, -1, and -2, in \textbf{Fig. S8} shows that there are more cases where the unoptimized energy is unexpectedly higher than the optimized energy. This may be a factor of the charged models never encountering pristine structures, as those are typically only utilized in neutral calculations and training sets. The charged models are only trained on partially optimized structures close to the ground state, or simply the ground state defect structures, making it such that q $\neq$ 0 ALIGNN predictions for pristine structures are less reliable than the neutral state predictions, even though the best charged models still have very low test RMSE values. \\

Finally, we examine the accuracy of predicting charge transition levels (CTLs) from ALIGNN compared to optimized DFT predictions. For any possible defect, a pristine unoptimized structure is created as described earlier and their DFE values are predicted at E$_{F}$ = 0 eV for q = 0, +2, +1, -1, and -2. Using these values, E$^{f}$ vs E$_{F}$ plots are produced based on the lowest energies and locations where the defect transitions from one stable charge state to another, referred to as $\epsilon$(q1/q2), are identified. This would effectively yield ALIGNN-unopt values for $\epsilon$(+2/+1), $\epsilon$(+1/0), $\epsilon$(0/-1), and $\epsilon$(-1/-2), which may then be compared with corresponding DFT-opt values available for datasets 1, 2, and 3. \textbf{Fig. \ref{fig:gnn_ctl} (b)} shows the ALIGNN-unopt $\epsilon$(+2/+1) plotted against the DFT-opt $\epsilon$(+2/+1), revealing substantial scatter and a lack of overall correlation. Similar behavior is observed for other CTLs as well, as shown in \textbf{Fig. S9}. This is not surprising, and indicates that the relative stability of different charged defects and thus their transitions are sensitive to defect geometry and would thus require some level of optimization. Thus, we conclude that although ALIGNN-unopt E$^{f}$(E$_{F}$=0) predictions within a threshold will provide some idea about the likelihood of formation of defects in a compound and its possible defect tolerance and dopability, exact CTLs will be harder to determine without optimization, which means ALIGNN-unopt alone cannot reveal the shallow or deep level nature of important low energy defects. \\

\subsection{ALIGNN-based defect structure optimization}

We employed our trained ALIGNN model to optimize a crystal containing point defects using an iterative distortion approach. This gradient-free optimization method entailed subjecting the initial defective crystal structure to a series of systematic atomic displacements. Each atom in the crystal was displaced based on a normal distribution, and the displacements were carefully adjusted to ensure no overall translation of the system. The magnitude of these displacements ranged from zero (representing no displacement) to a maximum value specified as a user input. The best ALIGNN models are applied to predict the E$^{f}$ of all generated crystals in all 5 q states. This procedure is iteratively repeated and the applied distortions are adjusted until the predicted DFE becomes as low as possible. This approach allows the efficient exploration of the energy landscape of the defective crystal, seeking configurations that approach the optimal structure. 
Another advantage of this gradient-free approach is that it does not rely on explicit training for atomic forces, which significantly increases the memory cost of GNNs at both training and prediction time.\\

For a demonstration of this procedure, we pick two defects in the neutral state, namely Re$_{Zn}$ and La$_{Zn}$, both in ZnO. The ALIGNN-based defective structure optimization scheme is pictured in \textbf{Fig. \ref{fig:alignn_opt}(a)} and results of the optimization procedure for Re$_{Zn}$ and La$_{Zn}$ in ZnO are presented in \textbf{Fig. \ref{fig:alignn_opt}(b)} and \textbf{(c)}, respectively. After applying 644 consecutive distortions on the Re$_{Zn}$ geometry, the ALIGNN-predicted DFE comes down from 5.93 eV to 5.31 eV, which is very close to the DFT-optimized DFE of 5.30 eV. For La$_{Zn}$, applying a total of 2407 distortions helps reduce the ALIGNN-predicted DFE from 5.23 eV to 3.35 eV, which is more than 1 eV higher than the DFT-optimized DFE of 2.20 eV. ALIGNN-optimization required approximately 12 minutes for Re$_{Zn}$ and 40 minutes for La$_{Zn}$ using a standard laptop to generate defect configurations and make instant ALIGNN predictions, which is a vast improvement on the $\sim$ 2500 core hours each DFT-optimization would require. Thus, this procedure will efficiently yield lower energy defective structures, though an exact match with DFT may not occur for every system. Finally, examining the lowest energy ALIGNN and DFT structures shows remarkable similarities between the two, as pictured in \textbf{Fig. S10}. \\

Next, we applied the same optimization scheme to 6 selected defects across 3 compounds, for different charge states, and plotted their ALIGNN-optimized E$^{f}$ vs E$_{F}$ in \textbf{Fig. \ref{fig:forme}} alongside the corresponding DFT-optimized plots. We find that ALIGNN-optimization produces defect formation energy plots for Co$_{Zn}$ and Si$_{Zn}$ in ZnS (II-VI semiconductor), Rh$_{i}$ and B$_{i}$ in AlP (III-V semiconductor), and Li$_{Si}$ and C$_{i}$ in Si (group IV semiconductor), that match almost perfectly with DFT DFEs for all cases. The most stable charge states, transition levels, and E$^{f}$ magnitudes are predicted to be very similar from both ALIGNN and DFT. \textbf{Fig. \ref{fig:op_ctl}} further shows the CTLs for a few selected defects plotted from ALIGNN and DFT, showing both ALIGNN-unopt and ALIGNN-opt values. It can be seen that ALIGNN-optimization brings down the DFT vs ALIGNN RMSE from 0.37 eV to 0.17 eV, which is a very respectable CTL prediction error that is far better than previous composition-based models and also very commensurate with the DFT-experiment RMSE of 0.21 eV established for the same chemical space. \\

Our results demonstrate the effectiveness of GNNs in guiding crystal structure optimization and highlight their potential for accelerating materials discovery and design. It should be noted that the geometry optimization process is not very clean, and there is no clear answer on how many distortions or atomic displacements must be applied for a given defect, although the unoptimized vs optimized energy visualization provides some insight into different types of defects. It is not easy to determine when to stop the optimization process, other than when the GNN-predicted energy does not reduce anymore, which does not negate the risk of local minima trapping. This process can also get expensive when applying for hundreds of thousands of defects, especially depending on the values of hyperparameters such as r$_{c}$; nevertheless, they are still meaningfully faster than complete DFT optimization. \\

\subsection{High-throughput screening of defects}

The best ALIGNN models were finally applied to predict the E$^{f}$ (E$_{F}$ = 0 eV) of all 12,474 possible single defects and impurities across the entire chemical space, in all 5 q states, at A-rich chemical potential conditions. These predictions were then used to generate E$^{f}$ vs E$_{F}$ plots for all defects spanning the experimental band gap of the semiconductor. To screen for potentially low energy defects, we look for E$^{f}$ becoming negative for any portion of the band gap, using a stringent threshold of 1 eV for neutral defects and 0 eV for charged defects. This yields a total of 1,281 defects that are very likely to be stable based on the ALIGNN-unopt predictions, though many more such low-energy defects may exist once ALIGNN-optimization is performed. \textbf{Table \ref{tab:screened}} contains a few examples of low-energy defects identified by ALIGNN. Prediction of DFE by ALIGNN for all possible 12,474 defects at the A-rich chemical potential conditions are added to the SI. This provides a great starting point for future studies and the quick identification of n-type or p-type dopants in any compound of interest. \\

\begin{table}[ht]
\centering
\begin{tabular}{|c|c|c|c|}
\hline
\rowcolor{gray!20} Semiconductor & Group & Defect \\
\hline
AlN & III-V & Mn\textsubscript{N}  \\
\hline
AlP & III-V & Cr\textsubscript{i,B}  \\
\hline
AlAs & III-V & V\textsubscript{i,B} \\
\hline
AlSb & III-V & N\textsubscript{Sb}  \\
\hline
BN & III-V & Al\textsubscript{B}  \\
\hline
BP & III-V & V\textsubscript{P}  \\
\hline
BAs & III-V & Cr\textsubscript{As}  \\
\hline
BSb & III-V & Cr\textsubscript{B}  \\
\hline
CdO & II-VI & Ti\textsubscript{Cd}  \\
\hline
CdS & II-VI & Mn\textsubscript{i,B}  \\
\hline
CdSe & II-VI & Cr\textsubscript{Cd}  \\
\hline
CdTe & II-VI & Sc\textsubscript{Cd}  \\
\hline
\end{tabular}
\caption{\label{tab:screened} Selected defects predicted to be low energy at A-rich conditions from ALIGNN models.}
\end{table}

\section{Conclusions}

In this work, we used state-of-the-art crystal graph-based neural networks to develop predictive models for defect formation energies in a chemical space of zincblende semiconductors, by learning from a substantial computational dataset containing optimized and partially optimized geometries. Three established GNN techniques, namely CGCNN, MEGNET, and ALIGNN, are tested in this work. The ALIGNN scheme shows the best prediction performance and is capable of high-accuracy prediction for native defects, impurities, complexes, and defects in alloys. While ALIGNN predictions made on hypothetical pristine defect structures deviate significantly from DFT-optimized defect formation energies, we demonstrate an ALIGNN-based defective geometry optimization approach which helps bridge the gap and bring down errors in predicting charge transition levels. The ALIGNN-unoptimized predictions made for the entire chemical space of $>$ 12,000 possible defects are released with this manuscript, along with necessary code and training data. We believe the DFT-GNN approach presented in this work will be highly consequential for screening across optoelectronically active point defects and functional dopants in technologically important semiconductors, even being applicable to all kinds of defect complexes. The next steps would involve developing a package to perform ALIGNN-based defect optimization, expanding the models to other semiconductor classes and higher levels of theories, and testing alternative ML and GNN approaches for further improvement. \\

\section*{Conflicts of Interest}

There are no conflicts to declare.

\section*{Data Availability}

All DFT data and GNN predictions are included with the SI as .csv files. All code can be found on Github: \href{https://github.com/msehabibur/defect_GNN_gen_1/tree/main/Codes}{Link}

\section*{Acknowledgements}

A.M.K. acknowledges support from the School of Materials Engineering at Purdue University under account number F.10023800.05.002, as well as support from Argonne National Laboratory under sub-contracts 21090590 and 22057223. A.M.K. also acknowledges insightful discussions with Dr. Mariana Bertoni at Arizona State University, Dr. Prashun Gorai at Colorado School of Mines, and Dr. Maria K.Y. Chan at Argonne National Laboratory. This research used resources from the National Energy Research Scientific Computing Center (NERSC), the Laboratory Computing Resource Center (LCRC) at Argonne National Laboratory, and the Rosen Center for Advanced Computing (RCAC) clusters at Purdue University. P.G. acknowledges IIT Madras for providing financial assistance through the "International Immersion Experience Travel Award" to visit Purdue University. Please note commercial software is identified to specify procedures. Such identification does not imply a recommendation by the National Institute of Standards and Technology (NIST).

\section*{Author Contributions}

A.M.K. conceived and planned the research project. DFT computations and GNN model training were performed by M.H.R., P.G., P.M., and A.M.K.; S.K.Y., G.P., B.D., and K.C. provided constant guidance and software support for the project and for editing the manuscript. M.H.R. and A.M.K. took the lead on writing and editing.

\bibliographystyle{rsc}
\bibliography{main}

\providecommand*{\mcitethebibliography}{\thebibliography}
\csname @ifundefined\endcsname{endmcitethebibliography}
{\let\endmcitethebibliography\endthebibliography}{}
\begin{mcitethebibliography}{61}
\providecommand*{\natexlab}[1]{#1}
\providecommand*{\mciteSetBstSublistMode}[1]{}
\providecommand*{\mciteSetBstMaxWidthForm}[2]{}
\providecommand*{\mciteBstWouldAddEndPuncttrue}
  {\def\EndOfBibitem{\unskip.}}
\providecommand*{\mciteBstWouldAddEndPunctfalse}
  {\let\EndOfBibitem\relax}
\providecommand*{\mciteSetBstMidEndSepPunct}[3]{}
\providecommand*{\mciteSetBstSublistLabelBeginEnd}[3]{}
\providecommand*{\EndOfBibitem}{}
\mciteSetBstSublistMode{f}
\mciteSetBstMaxWidthForm{subitem}
{(\emph{\alph{mcitesubitemcount}})}
\mciteSetBstSublistLabelBeginEnd{\mcitemaxwidthsubitemform\space}
{\relax}{\relax}

\bibitem[Ganose \emph{et~al.}(2022)Ganose, Scanlon, Walsh, and Hoye]{Ganose_Scanlon_2022}
A.~M. Ganose, D.~O. Scanlon, A.~Walsh and R.~L.~Z. Hoye, \emph{Nature Communications}, 2022, \textbf{13}, 4715\relax
\mciteBstWouldAddEndPuncttrue
\mciteSetBstMidEndSepPunct{\mcitedefaultmidpunct}
{\mcitedefaultendpunct}{\mcitedefaultseppunct}\relax
\EndOfBibitem
\bibitem[CHI(August 2022)]{CHIPS}
\emph{The White House}, August 2022\relax
\mciteBstWouldAddEndPuncttrue
\mciteSetBstMidEndSepPunct{\mcitedefaultmidpunct}
{\mcitedefaultendpunct}{\mcitedefaultseppunct}\relax
\EndOfBibitem
\bibitem[Zwerver \emph{et~al.}(2022)Zwerver, Krähenmann, and et~al.]{J_2022}
A.~M.~J. Zwerver, T.~Krähenmann and et~al., \emph{Nature Electronics}, 2022, \textbf{5}, 184–190\relax
\mciteBstWouldAddEndPuncttrue
\mciteSetBstMidEndSepPunct{\mcitedefaultmidpunct}
{\mcitedefaultendpunct}{\mcitedefaultseppunct}\relax
\EndOfBibitem
\bibitem[Chen \emph{et~al.}(2013)Chen, Walsh, Gong, and Wei]{Chen_Walsh_Gong_Wei_2013}
S.~Chen, A.~Walsh, X.-G. Gong and S.-H. Wei, \emph{Advanced Materials}, 2013, \textbf{25}, 1522–1539\relax
\mciteBstWouldAddEndPuncttrue
\mciteSetBstMidEndSepPunct{\mcitedefaultmidpunct}
{\mcitedefaultendpunct}{\mcitedefaultseppunct}\relax
\EndOfBibitem
\bibitem[Mannodi-Kanakkithodi(2023)]{Mannodi-Kanakkithodi_2023}
A.~Mannodi-Kanakkithodi, \emph{The devil is in the defects - Nature Physics}, 2023, \url{https://www.nature.com/articles/s41567-023-02049-9}\relax
\mciteBstWouldAddEndPuncttrue
\mciteSetBstMidEndSepPunct{\mcitedefaultmidpunct}
{\mcitedefaultendpunct}{\mcitedefaultseppunct}\relax
\EndOfBibitem
\bibitem[Mannodi-Kanakkithodi \emph{et~al.}(2020)Mannodi-Kanakkithodi, Park, Martinson, and Chan]{Defect_Energetics_JPCC}
A.~Mannodi-Kanakkithodi, J.-S. Park, A.~B.~F. Martinson and M.~K.~Y. Chan, \emph{The Journal of Physical Chemistry C}, 2020, \textbf{124}, 16729--16738\relax
\mciteBstWouldAddEndPuncttrue
\mciteSetBstMidEndSepPunct{\mcitedefaultmidpunct}
{\mcitedefaultendpunct}{\mcitedefaultseppunct}\relax
\EndOfBibitem
\bibitem[Mannodi-Kanakkithodi \emph{et~al.}(2019)Mannodi-Kanakkithodi, Park, Jeon, Cao, Gosztola, Martinson, and Chan]{Chemistry_of_MSE}
A.~Mannodi-Kanakkithodi, J.-S. Park, N.~Jeon, D.~H. Cao, D.~J. Gosztola, A.~B.~F. Martinson and M.~K.~Y. Chan, \emph{Chemistry of Materials}, 2019, \textbf{31}, 3599–3612\relax
\mciteBstWouldAddEndPuncttrue
\mciteSetBstMidEndSepPunct{\mcitedefaultmidpunct}
{\mcitedefaultendpunct}{\mcitedefaultseppunct}\relax
\EndOfBibitem
\bibitem[Broberg \emph{et~al.}(2023)Broberg, Bystrom, and et~al]{Broberg_Bystrom}
D.~Broberg, K.~Bystrom and et~al, \emph{npj Computational Materials}, 2023, \textbf{9}, 72\relax
\mciteBstWouldAddEndPuncttrue
\mciteSetBstMidEndSepPunct{\mcitedefaultmidpunct}
{\mcitedefaultendpunct}{\mcitedefaultseppunct}\relax
\EndOfBibitem
\bibitem[Buckeridge(2019)]{Buckeridge_2019}
J.~Buckeridge, \emph{Computer Physics Communications}, 2019, \textbf{244}, 329–342\relax
\mciteBstWouldAddEndPuncttrue
\mciteSetBstMidEndSepPunct{\mcitedefaultmidpunct}
{\mcitedefaultendpunct}{\mcitedefaultseppunct}\relax
\EndOfBibitem
\bibitem[Turiansky \emph{et~al.}(2021)Turiansky, Alkauskas, Engel, Kresse, Wickramaratne, Shen, Dreyer, and Van~de Walle]{Turiansky}
M.~E. Turiansky, A.~Alkauskas, M.~Engel, G.~Kresse, D.~Wickramaratne, J.-X. Shen, C.~E. Dreyer and C.~G. Van~de Walle, \emph{Computer Physics Communications}, 2021, \textbf{267}, 108056\relax
\mciteBstWouldAddEndPuncttrue
\mciteSetBstMidEndSepPunct{\mcitedefaultmidpunct}
{\mcitedefaultendpunct}{\mcitedefaultseppunct}\relax
\EndOfBibitem
\bibitem[Lannoo and Bourgoin(1981)]{Lannoo1981}
M.~Lannoo and J.~Bourgoin, in \emph{Defect Migration and Diffusion}, Springer Berlin Heidelberg, Berlin, Heidelberg, 1981, pp. 219--244\relax
\mciteBstWouldAddEndPuncttrue
\mciteSetBstMidEndSepPunct{\mcitedefaultmidpunct}
{\mcitedefaultendpunct}{\mcitedefaultseppunct}\relax
\EndOfBibitem
\bibitem[Mannodi-Kanakkithodi \emph{et~al.}(2022)Mannodi-Kanakkithodi, Xiang, Jacoby, Biegaj, Dunham, Gamelin, and Chan]{Mannodi-Kanakkithodi2022-ck}
A.~Mannodi-Kanakkithodi, X.~Xiang, L.~Jacoby, R.~Biegaj, S.~T. Dunham, D.~R. Gamelin and M.~K.~Y. Chan, \emph{Patterns (N. Y.)}, 2022, \textbf{3}, 100450\relax
\mciteBstWouldAddEndPuncttrue
\mciteSetBstMidEndSepPunct{\mcitedefaultmidpunct}
{\mcitedefaultendpunct}{\mcitedefaultseppunct}\relax
\EndOfBibitem
\bibitem[Chen \emph{et~al.}(2010)Chen, Yang, Gong, Walsh, and Wei]{Chen_Yang_2010}
S.~Chen, J.-H. Yang, X.~G. Gong, A.~Walsh and S.-H. Wei, \emph{Physical Review B}, 2010, \textbf{81}, 245204\relax
\mciteBstWouldAddEndPuncttrue
\mciteSetBstMidEndSepPunct{\mcitedefaultmidpunct}
{\mcitedefaultendpunct}{\mcitedefaultseppunct}\relax
\EndOfBibitem
\bibitem[Queisser and Haller(1998)]{Queisser_Haller_1998}
H.~J. Queisser and E.~E. Haller, \emph{Science}, 1998, \textbf{281}, 945–950\relax
\mciteBstWouldAddEndPuncttrue
\mciteSetBstMidEndSepPunct{\mcitedefaultmidpunct}
{\mcitedefaultendpunct}{\mcitedefaultseppunct}\relax
\EndOfBibitem
\bibitem[Liu \emph{et~al.}(2020)Liu, Na, Tian, Yu, Li, and Zhang]{Liu_2020}
Z.~Liu, G.~Na, F.~Tian, L.~Yu, J.~Li and L.~Zhang, \emph{InfoMat}, 2020, \textbf{2}, 879–904\relax
\mciteBstWouldAddEndPuncttrue
\mciteSetBstMidEndSepPunct{\mcitedefaultmidpunct}
{\mcitedefaultendpunct}{\mcitedefaultseppunct}\relax
\EndOfBibitem
\bibitem[Nayak \emph{et~al.}(2019)Nayak, Mahesh, Snaith, and Cahen]{Nayak_2019}
P.~K. Nayak, S.~Mahesh, H.~J. Snaith and D.~Cahen, \emph{Nature Reviews Materials}, 2019, \textbf{4}, 269–285\relax
\mciteBstWouldAddEndPuncttrue
\mciteSetBstMidEndSepPunct{\mcitedefaultmidpunct}
{\mcitedefaultendpunct}{\mcitedefaultseppunct}\relax
\EndOfBibitem
\bibitem[Sivathanu \emph{et~al.}(2021)Sivathanu, R, and Lenka]{Sivathanu_R_Lenka_2021}
V.~Sivathanu, T.~R and T.~R. Lenka, \emph{International Journal of Energy Research}, 2021, \textbf{45}, 10527–10537\relax
\mciteBstWouldAddEndPuncttrue
\mciteSetBstMidEndSepPunct{\mcitedefaultmidpunct}
{\mcitedefaultendpunct}{\mcitedefaultseppunct}\relax
\EndOfBibitem
\bibitem[Kazeev and et~al.(2023)]{Kazeev_Al}
N.~Kazeev and et~al., \emph{npj Computational Materials}, 2023, \textbf{9}, 113\relax
\mciteBstWouldAddEndPuncttrue
\mciteSetBstMidEndSepPunct{\mcitedefaultmidpunct}
{\mcitedefaultendpunct}{\mcitedefaultseppunct}\relax
\EndOfBibitem
\bibitem[Gorai \emph{et~al.}(2019)Gorai, McKinney, Haegel, Zakutayev, and Stevanovic]{Gorai}
P.~Gorai, R.~W. McKinney, N.~M. Haegel, A.~Zakutayev and V.~Stevanovic, \emph{Energy Environ. Sci.}, 2019, \textbf{12}, 3338--3347\relax
\mciteBstWouldAddEndPuncttrue
\mciteSetBstMidEndSepPunct{\mcitedefaultmidpunct}
{\mcitedefaultendpunct}{\mcitedefaultseppunct}\relax
\EndOfBibitem
\bibitem[Schirhagl \emph{et~al.}(2014)Schirhagl, Chang, Loretz, and Degen]{nv}
R.~Schirhagl, K.~Chang, M.~Loretz and C.~L. Degen, \emph{Annual Review of Physical Chemistry}, 2014, \textbf{65}, 83--105\relax
\mciteBstWouldAddEndPuncttrue
\mciteSetBstMidEndSepPunct{\mcitedefaultmidpunct}
{\mcitedefaultendpunct}{\mcitedefaultseppunct}\relax
\EndOfBibitem
\bibitem[Srivastava and et~al.(2023)]{Srivastava_Ranjan}
S.~Srivastava and et~al., \emph{Communications Materials}, 2023, \textbf{4}, 52\relax
\mciteBstWouldAddEndPuncttrue
\mciteSetBstMidEndSepPunct{\mcitedefaultmidpunct}
{\mcitedefaultendpunct}{\mcitedefaultseppunct}\relax
\EndOfBibitem
\bibitem[Kim \emph{et~al.}(2019)Kim, Park, Hood, and Walsh]{Kim_Park_Hood_Walsh_2019}
S.~Kim, J.-S. Park, S.~Hood and A.~Walsh, \emph{Journal of Materials Chemistry A}, 2019, \textbf{7}, 2686–2693\relax
\mciteBstWouldAddEndPuncttrue
\mciteSetBstMidEndSepPunct{\mcitedefaultmidpunct}
{\mcitedefaultendpunct}{\mcitedefaultseppunct}\relax
\EndOfBibitem
\bibitem[Rahman \emph{et~al.}(2023)Rahman, Yang, Sun, and Mannodi-Kanakkithodi]{Rahman_Mannodi-Kanakkithodi_2023}
M.~H. Rahman, J.~Yang, Y.~Sun and A.~Mannodi-Kanakkithodi, \emph{Surfaces and Interfaces}, 2023, \textbf{39}, 102960\relax
\mciteBstWouldAddEndPuncttrue
\mciteSetBstMidEndSepPunct{\mcitedefaultmidpunct}
{\mcitedefaultendpunct}{\mcitedefaultseppunct}\relax
\EndOfBibitem
\bibitem[Mannodi-Kanakkithodi \emph{et~al.}(2020)Mannodi-Kanakkithodi, Toriyama, Sen, Davis, Klie, and Chan]{Mannodi_Toriyama_2020}
A.~Mannodi-Kanakkithodi, M.~Y. Toriyama, F.~G. Sen, M.~J. Davis, R.~F. Klie and M.~K.~Y. Chan, \emph{npj Computational Materials}, 2020, \textbf{6}, 39\relax
\mciteBstWouldAddEndPuncttrue
\mciteSetBstMidEndSepPunct{\mcitedefaultmidpunct}
{\mcitedefaultendpunct}{\mcitedefaultseppunct}\relax
\EndOfBibitem
\bibitem[Witman \emph{et~al.}(2023)Witman, Goyal, Ogitsu, McDaniel, and Lany]{Witman_2023}
M.~D. Witman, A.~Goyal, T.~Ogitsu, A.~H. McDaniel and S.~Lany, \emph{Nature Computational Science}, 2023\relax
\mciteBstWouldAddEndPuncttrue
\mciteSetBstMidEndSepPunct{\mcitedefaultmidpunct}
{\mcitedefaultendpunct}{\mcitedefaultseppunct}\relax
\EndOfBibitem
\bibitem[Shimizu \emph{et~al.}(2022)Shimizu, Dou, Arguelles, Moriya, Minamitani, and Watanabe]{def_ml1}
K.~Shimizu, Y.~Dou, E.~F. Arguelles, T.~Moriya, E.~Minamitani and S.~Watanabe, \emph{Phys. Rev. B}, 2022, \textbf{106}, 054108\relax
\mciteBstWouldAddEndPuncttrue
\mciteSetBstMidEndSepPunct{\mcitedefaultmidpunct}
{\mcitedefaultendpunct}{\mcitedefaultseppunct}\relax
\EndOfBibitem
\bibitem[Frey \emph{et~al.}(2020)Frey, Akinwande, Jariwala, and Shenoy]{def_ml2}
N.~C. Frey, D.~Akinwande, D.~Jariwala and V.~B. Shenoy, \emph{ACS Nano}, 2020, \textbf{14}, 13406--13417\relax
\mciteBstWouldAddEndPuncttrue
\mciteSetBstMidEndSepPunct{\mcitedefaultmidpunct}
{\mcitedefaultendpunct}{\mcitedefaultseppunct}\relax
\EndOfBibitem
\bibitem[Bartel \emph{et~al.}(2020)Bartel, Trewartha, Wang, Dunn, Jain, and Ceder]{Bartel}
C.~J. Bartel, A.~Trewartha, Q.~Wang, A.~Dunn, A.~Jain and G.~Ceder, \emph{npj Computational Materials}, 2020, \textbf{6}, 97\relax
\mciteBstWouldAddEndPuncttrue
\mciteSetBstMidEndSepPunct{\mcitedefaultmidpunct}
{\mcitedefaultendpunct}{\mcitedefaultseppunct}\relax
\EndOfBibitem
\bibitem[Fung \emph{et~al.}(2021)Fung, Zhang, Juarez, and Sumpter]{Fung}
V.~Fung, J.~Zhang, E.~Juarez and B.~G. Sumpter, \emph{npj Computational Materials}, 2021, \textbf{7}, 84\relax
\mciteBstWouldAddEndPuncttrue
\mciteSetBstMidEndSepPunct{\mcitedefaultmidpunct}
{\mcitedefaultendpunct}{\mcitedefaultseppunct}\relax
\EndOfBibitem
\bibitem[Chen and Ong(2022)]{Chen_Ong_2022}
C.~Chen and S.~P. Ong, \emph{Nature Computational Science}, 2022, \textbf{2}, 718–728\relax
\mciteBstWouldAddEndPuncttrue
\mciteSetBstMidEndSepPunct{\mcitedefaultmidpunct}
{\mcitedefaultendpunct}{\mcitedefaultseppunct}\relax
\EndOfBibitem
\bibitem[Cheng \emph{et~al.}(2021)Cheng, Zhang, and Dong]{Cheng}
J.~Cheng, C.~Zhang and L.~Dong, \emph{Communications Materials}, 2021, \textbf{2}, 92\relax
\mciteBstWouldAddEndPuncttrue
\mciteSetBstMidEndSepPunct{\mcitedefaultmidpunct}
{\mcitedefaultendpunct}{\mcitedefaultseppunct}\relax
\EndOfBibitem
\bibitem[Park \emph{et~al.}(2021)Park, Kornbluth, Vandermause, Wolverton, Kozinsky, and Mailoa]{Park}
C.~W. Park, M.~Kornbluth, J.~Vandermause, C.~Wolverton, B.~Kozinsky and J.~P. Mailoa, \emph{npj Computational Materials}, 2021, \textbf{7}, 73\relax
\mciteBstWouldAddEndPuncttrue
\mciteSetBstMidEndSepPunct{\mcitedefaultmidpunct}
{\mcitedefaultendpunct}{\mcitedefaultseppunct}\relax
\EndOfBibitem
\bibitem[Bhattacharya \emph{et~al.}(2023)Bhattacharya, Timokhin, Chatterjee, Yang, and Mishchenko]{Bhattacharya_2023}
A.~Bhattacharya, I.~Timokhin, R.~Chatterjee, Q.~Yang and A.~Mishchenko, \emph{npj Computational Materials}, 2023, \textbf{9}, 101\relax
\mciteBstWouldAddEndPuncttrue
\mciteSetBstMidEndSepPunct{\mcitedefaultmidpunct}
{\mcitedefaultendpunct}{\mcitedefaultseppunct}\relax
\EndOfBibitem
\bibitem[Kilgour \emph{et~al.}(2023)Kilgour, Rogal, and Tuckerman]{Kilgour_Rogal_Tuckerman_2023}
M.~Kilgour, J.~Rogal and M.~Tuckerman, \emph{Journal of Chemical Theory and Computation}, 2023, \textbf{19}, 4743–4756\relax
\mciteBstWouldAddEndPuncttrue
\mciteSetBstMidEndSepPunct{\mcitedefaultmidpunct}
{\mcitedefaultendpunct}{\mcitedefaultseppunct}\relax
\EndOfBibitem
\bibitem[Bapst \emph{et~al.}(2020)Bapst, Keck, Grabska-Barwińska, Donner, Cubuk, Schoenholz, Obika, Nelson, Back, Hassabis, and Kohli]{Bapst_Keck}
V.~Bapst, T.~Keck, A.~Grabska-Barwińska, C.~Donner, E.~D. Cubuk, S.~S. Schoenholz, A.~Obika, A.~W.~R. Nelson, T.~Back, D.~Hassabis and P.~Kohli, \emph{Nature Physics}, 2020, \textbf{16}, 448–454\relax
\mciteBstWouldAddEndPuncttrue
\mciteSetBstMidEndSepPunct{\mcitedefaultmidpunct}
{\mcitedefaultendpunct}{\mcitedefaultseppunct}\relax
\EndOfBibitem
\bibitem[Xie and Grossman(2018)]{CGCNN}
T.~Xie and J.~C. Grossman, \emph{Phys. Rev. Lett.}, 2018, \textbf{120}, 145301\relax
\mciteBstWouldAddEndPuncttrue
\mciteSetBstMidEndSepPunct{\mcitedefaultmidpunct}
{\mcitedefaultendpunct}{\mcitedefaultseppunct}\relax
\EndOfBibitem
\bibitem[Chen \emph{et~al.}(2019)Chen, Ye, Zuo, Zheng, and Ong]{MEGNET}
C.~Chen, W.~Ye, Y.~Zuo, C.~Zheng and S.~P. Ong, \emph{Chemistry of Materials}, 2019, \textbf{31}, 3564--3572\relax
\mciteBstWouldAddEndPuncttrue
\mciteSetBstMidEndSepPunct{\mcitedefaultmidpunct}
{\mcitedefaultendpunct}{\mcitedefaultseppunct}\relax
\EndOfBibitem
\bibitem[Choudhary and DeCost(2022)]{ALIGNN}
K.~Choudhary and B.~DeCost, \emph{npj Computational Materials}, 2022, \textbf{8}, 221\relax
\mciteBstWouldAddEndPuncttrue
\mciteSetBstMidEndSepPunct{\mcitedefaultmidpunct}
{\mcitedefaultendpunct}{\mcitedefaultseppunct}\relax
\EndOfBibitem
\bibitem[Rahman \emph{et~al.}(2023)Rahman, Rojsatien, Bertoni, Chan, and Mannodi-Kanakkithodi]{CdSeTe_def}
H.~Rahman, S.~Rojsatien, M.~Bertoni, M.~Chan and A.~Mannodi-Kanakkithodi, \emph{In Preparation}, 2023\relax
\mciteBstWouldAddEndPuncttrue
\mciteSetBstMidEndSepPunct{\mcitedefaultmidpunct}
{\mcitedefaultendpunct}{\mcitedefaultseppunct}\relax
\EndOfBibitem
\bibitem[Heyd \emph{et~al.}(2003)Heyd, Scuseria, and Ernzerhof]{hse}
J.~Heyd, G.~E. Scuseria and M.~Ernzerhof, \emph{The Journal of Chemical Physics}, 2003, \textbf{118}, 8207--8215\relax
\mciteBstWouldAddEndPuncttrue
\mciteSetBstMidEndSepPunct{\mcitedefaultmidpunct}
{\mcitedefaultendpunct}{\mcitedefaultseppunct}\relax
\EndOfBibitem
\bibitem[Vona \emph{et~al.}(2022)Vona, Nabok, and Draxl]{HP_hse}
C.~Vona, D.~Nabok and C.~Draxl, \emph{Advanced Theory and Simulations}, 2022, \textbf{5}, 2100496\relax
\mciteBstWouldAddEndPuncttrue
\mciteSetBstMidEndSepPunct{\mcitedefaultmidpunct}
{\mcitedefaultendpunct}{\mcitedefaultseppunct}\relax
\EndOfBibitem
\bibitem[Vona \emph{et~al.}(2022)Vona, Nabok, and Draxl]{gw}
C.~Vona, D.~Nabok and C.~Draxl, \emph{Advanced Theory and Simulations}, 2022, \textbf{5}, 2100496\relax
\mciteBstWouldAddEndPuncttrue
\mciteSetBstMidEndSepPunct{\mcitedefaultmidpunct}
{\mcitedefaultendpunct}{\mcitedefaultseppunct}\relax
\EndOfBibitem
\bibitem[Polak \emph{et~al.}(2022)Polak, Jacobs, Mannodi-Kanakkithodi, Chan, and Morgan]{mba}
M.~P. Polak, R.~Jacobs, A.~Mannodi-Kanakkithodi, M.~K.~Y. Chan and D.~Morgan, \emph{The Journal of Chemical Physics}, 2022, \textbf{156}, 114110\relax
\mciteBstWouldAddEndPuncttrue
\mciteSetBstMidEndSepPunct{\mcitedefaultmidpunct}
{\mcitedefaultendpunct}{\mcitedefaultseppunct}\relax
\EndOfBibitem
\bibitem[Toriyama \emph{et~al.}(2021)Toriyama, Qu, Snyder, and Gorai]{gw_def}
M.~Y. Toriyama, J.~Qu, G.~J. Snyder and P.~Gorai, \emph{J. Mater. Chem. A}, 2021, \textbf{9}, 20685--20694\relax
\mciteBstWouldAddEndPuncttrue
\mciteSetBstMidEndSepPunct{\mcitedefaultmidpunct}
{\mcitedefaultendpunct}{\mcitedefaultseppunct}\relax
\EndOfBibitem
\bibitem[Kresse and Furthm\"uller(1996)]{vasp}
G.~Kresse and J.~Furthm\"uller, \emph{Phys. Rev. B}, 1996, \textbf{54}, 11169--11186\relax
\mciteBstWouldAddEndPuncttrue
\mciteSetBstMidEndSepPunct{\mcitedefaultmidpunct}
{\mcitedefaultendpunct}{\mcitedefaultseppunct}\relax
\EndOfBibitem
\bibitem[Li \emph{et~al.}(2017)Li, Poplawsky, Yan, and Pennycook]{Li2017-ez}
C.~Li, J.~Poplawsky, Y.~Yan and S.~J. Pennycook, \emph{Mater. Sci. Semicond. Process.}, 2017, \textbf{65}, 64--76\relax
\mciteBstWouldAddEndPuncttrue
\mciteSetBstMidEndSepPunct{\mcitedefaultmidpunct}
{\mcitedefaultendpunct}{\mcitedefaultseppunct}\relax
\EndOfBibitem
\bibitem[Poplawsky \emph{et~al.}(2016)Poplawsky, Guo, Paudel, Ng, More, Leonard, and Yan]{Poplawsky2016-hq}
J.~D. Poplawsky, W.~Guo, N.~Paudel, A.~Ng, K.~More, D.~Leonard and Y.~Yan, \emph{Nat. Commun.}, 2016, \textbf{7}, year\relax
\mciteBstWouldAddEndPuncttrue
\mciteSetBstMidEndSepPunct{\mcitedefaultmidpunct}
{\mcitedefaultendpunct}{\mcitedefaultseppunct}\relax
\EndOfBibitem
\bibitem[Gloeckler \emph{et~al.}(2013)Gloeckler, Sankin, and Zhao]{CdTe_1}
M.~Gloeckler, I.~Sankin and Z.~Zhao, \emph{IEEE Journal of Photovoltaics}, 2013, \textbf{3}, 1389--1393\relax
\mciteBstWouldAddEndPuncttrue
\mciteSetBstMidEndSepPunct{\mcitedefaultmidpunct}
{\mcitedefaultendpunct}{\mcitedefaultseppunct}\relax
\EndOfBibitem
\bibitem[Nideep \emph{et~al.}(2020)Nideep, Ramya, and Kailasnath]{CdTe_2}
T.~Nideep, M.~Ramya and M.~Kailasnath, \emph{Superlattices and Microstructures}, 2020, \textbf{141}, 106477\relax
\mciteBstWouldAddEndPuncttrue
\mciteSetBstMidEndSepPunct{\mcitedefaultmidpunct}
{\mcitedefaultendpunct}{\mcitedefaultseppunct}\relax
\EndOfBibitem
\bibitem[Yang \emph{et~al.}(2023)Yang, Long, Zheng, Wang, Zhou, Xie, Li, Zhang, Hao, Karazhanov, Zeng, and Feng]{CdTe_3}
X.~Yang, Y.~Long, Y.~Zheng, J.~Wang, B.~Zhou, S.~Xie, B.~Li, J.~Zhang, X.~Hao, S.~Karazhanov, G.~Zeng and L.~Feng, \emph{Materials Science in Semiconductor Processing}, 2023, \textbf{156}, 107267\relax
\mciteBstWouldAddEndPuncttrue
\mciteSetBstMidEndSepPunct{\mcitedefaultmidpunct}
{\mcitedefaultendpunct}{\mcitedefaultseppunct}\relax
\EndOfBibitem
\bibitem[Wickramaratne \emph{et~al.}(2018)Wickramaratne, Dreyer, Monserrat, Shen, Lyons, Alkauskas, and Van~de Walle]{Wickramaratne}
D.~Wickramaratne, C.~E. Dreyer, B.~Monserrat, J.-X. Shen, J.~L. Lyons, A.~Alkauskas and C.~G. Van~de Walle, \emph{Applied Physics Letters}, 2018, \textbf{113}, 192106\relax
\mciteBstWouldAddEndPuncttrue
\mciteSetBstMidEndSepPunct{\mcitedefaultmidpunct}
{\mcitedefaultendpunct}{\mcitedefaultseppunct}\relax
\EndOfBibitem
\bibitem[Lee and Asahi(2021)]{Lee_Asahi}
J.~Lee and R.~Asahi, \emph{Computational Materials Science}, 2021, \textbf{190}, 110314\relax
\mciteBstWouldAddEndPuncttrue
\mciteSetBstMidEndSepPunct{\mcitedefaultmidpunct}
{\mcitedefaultendpunct}{\mcitedefaultseppunct}\relax
\EndOfBibitem
\bibitem[Kipf and Welling(2016)]{Kipf_Welling_2016}
T.~N. Kipf and M.~Welling, \emph{Semi-Supervised Classification with Graph Convolutional Networks}, 2016, \url{https://arxiv.org/abs/1609.02907v4}\relax
\mciteBstWouldAddEndPuncttrue
\mciteSetBstMidEndSepPunct{\mcitedefaultmidpunct}
{\mcitedefaultendpunct}{\mcitedefaultseppunct}\relax
\EndOfBibitem
\bibitem[Zhou \emph{et~al.}(2018)Zhou, Cui, Hu, Zhang, Yang, Liu, Wang, Li, and Sun]{Zhou_Cui}
J.~Zhou, G.~Cui, S.~Hu, Z.~Zhang, C.~Yang, Z.~Liu, L.~Wang, C.~Li and M.~Sun, \emph{Graph Neural Networks: A Review of Methods and Applications}, 2018, \url{https://arxiv.org/abs/1812.08434v6}\relax
\mciteBstWouldAddEndPuncttrue
\mciteSetBstMidEndSepPunct{\mcitedefaultmidpunct}
{\mcitedefaultendpunct}{\mcitedefaultseppunct}\relax
\EndOfBibitem
\bibitem[Chen \emph{et~al.}(2017)Chen, Li, and Bruna]{Chen_Li_Bruna_2017}
Z.~Chen, X.~Li and J.~Bruna, \emph{Supervised Community Detection with Line Graph Neural Networks}, 2017, \url{https://arxiv.org/abs/1705.08415v6}\relax
\mciteBstWouldAddEndPuncttrue
\mciteSetBstMidEndSepPunct{\mcitedefaultmidpunct}
{\mcitedefaultendpunct}{\mcitedefaultseppunct}\relax
\EndOfBibitem
\bibitem[Kingma and Ba(2014)]{kingma2014adam}
D.~P. Kingma and J.~Ba, \emph{arXiv preprint arXiv:1412.6980}, 2014\relax
\mciteBstWouldAddEndPuncttrue
\mciteSetBstMidEndSepPunct{\mcitedefaultmidpunct}
{\mcitedefaultendpunct}{\mcitedefaultseppunct}\relax
\EndOfBibitem
\bibitem[Pandey \emph{et~al.}(2021)Pandey, Qu, Stevanović, {St. John}, and Gorai]{ube1}
S.~Pandey, J.~Qu, V.~Stevanović, P.~{St. John} and P.~Gorai, \emph{Patterns}, 2021, \textbf{2}, 100361\relax
\mciteBstWouldAddEndPuncttrue
\mciteSetBstMidEndSepPunct{\mcitedefaultmidpunct}
{\mcitedefaultendpunct}{\mcitedefaultseppunct}\relax
\EndOfBibitem
\bibitem[Law \emph{et~al.}(2023)Law, Pandey, Gorai, and St.~John]{ube2}
J.~N. Law, S.~Pandey, P.~Gorai and P.~C. St.~John, \emph{JACS Au}, 2023, \textbf{3}, 113--123\relax
\mciteBstWouldAddEndPuncttrue
\mciteSetBstMidEndSepPunct{\mcitedefaultmidpunct}
{\mcitedefaultendpunct}{\mcitedefaultseppunct}\relax
\EndOfBibitem
\bibitem[Choudhary \emph{et~al.}(2023)Choudhary, DeCost, Major, Butler, Thiyagalingam, and Tavazza]{Choudhary_DeCost}
K.~Choudhary, B.~DeCost, L.~Major, K.~Butler, J.~Thiyagalingam and F.~Tavazza, \emph{Digital Discovery}, 2023, \textbf{2}, 346–355\relax
\mciteBstWouldAddEndPuncttrue
\mciteSetBstMidEndSepPunct{\mcitedefaultmidpunct}
{\mcitedefaultendpunct}{\mcitedefaultseppunct}\relax
\EndOfBibitem
\bibitem[Schütt \emph{et~al.}(2018)Schütt, Sauceda, Kindermans, Tkatchenko, and Müller]{SchNet}
K.~T. Schütt, H.~E. Sauceda, P.-J. Kindermans, A.~Tkatchenko and K.-R. Müller, \emph{The Journal of Chemical Physics}, 2018, \textbf{148}, 241722\relax
\mciteBstWouldAddEndPuncttrue
\mciteSetBstMidEndSepPunct{\mcitedefaultmidpunct}
{\mcitedefaultendpunct}{\mcitedefaultseppunct}\relax
\EndOfBibitem
\bibitem[Chen and Ong(2022)]{m3gnet}
C.~Chen and S.~P. Ong, \emph{Nature Computational Science}, 2022, \textbf{2}, 718--728\relax
\mciteBstWouldAddEndPuncttrue
\mciteSetBstMidEndSepPunct{\mcitedefaultmidpunct}
{\mcitedefaultendpunct}{\mcitedefaultseppunct}\relax
\EndOfBibitem
\end{mcitethebibliography}

\clearpage
\newpage
\pagenumbering{gobble}
\thispagestyle{empty} 

\onecolumn

\setcounter{figure}{0}   
\setcounter{table}{0} 
\renewcommand{\thetable}{S\Roman{table}} 
\renewcommand\thefigure{S\arabic{figure}}

\begin{center}
\vspace{0.5cm}
\Large
\textbf{Supplemental material to "Accelerating Defect Prediction in Semiconductors Using Graph Neural Networks"\\}
\vspace{0.5cm}
\large

 \noindent\large{Md. Habibur Rahman\textsuperscript{a}, Prince Gollapalli\textsuperscript{b}, Panayotis Manganaris\textsuperscript{a}, Satyesh Kumar Yadav\textsuperscript{b, c}, Ghanshyam Pilania\textsuperscript{d}, Brian DeCost\textsuperscript{e}, Kamal Choudhary\textsuperscript{e}, and Arun Mannodi-Kanakkithodi\textsuperscript{a}}\\

\vspace{0.3cm}

\normalsize

  \textsuperscript{a}School of Materials Engineering, Purdue University, West Lafayette, IN 47907, USA; E-mail: amannodi@purdue.edu

  \textsuperscript{b}Department of Metallurgical and Materials Engineering, Indian Institute of Technology (IIT) Madras, Chennai 600036, India

   \textsuperscript{c} Centre for Atomistic Modelling and Materials Design, Indian Institute of Technology (IIT) Madras, Chennai, 600036, India
  
   \textsuperscript{d}GE Research, Schenectady, NY, 12309, USA

   \textsuperscript{e}Materials Measurement Laboratory, National Institute of Standards and Technology, Gaithersburg, MD, 20899, USA


\end{center}

\footnote{
\textsuperscript{a}amannodi@purdue.edu\hspace{0.3cm}}

\begin{figure*}[htb]
  \centering
  \includegraphics[width=0.9\linewidth, height=12cm]{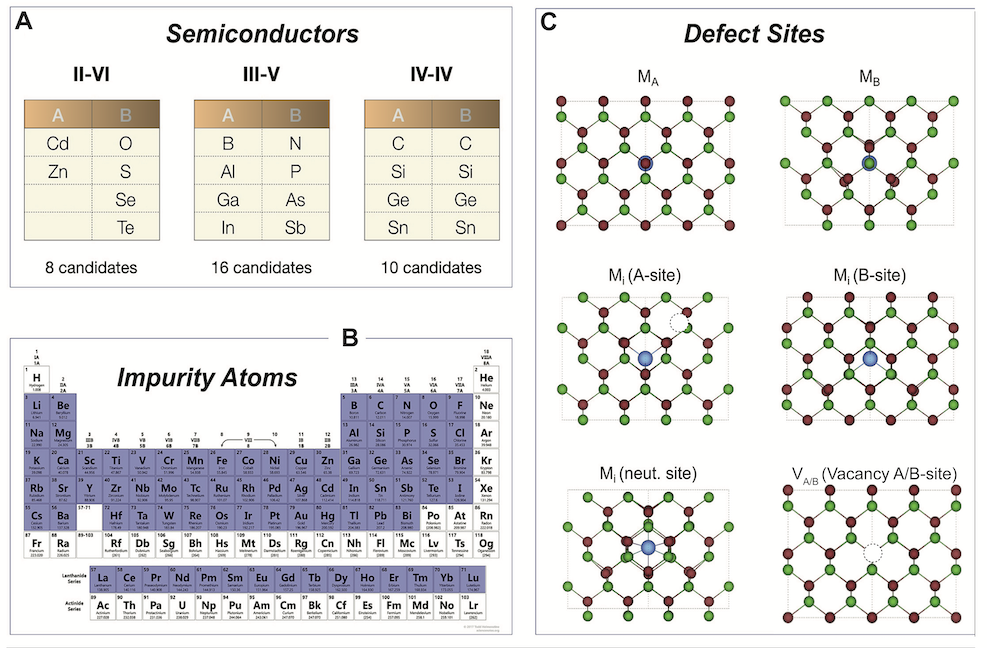}
  \caption{\label{fig:S1} Visualization of the semiconductor-defect chemical space in terms of (a) all group IV, III-V, and II-VI semiconductors, (b) all possible defect/impurity atoms, and (c) possible defect sites considered in the zincblende structure.}
  \vspace{0.5cm}
\end{figure*}

\begin{figure*}[htb]
  \centering
  \includegraphics[width=1.0\linewidth, height=7cm]{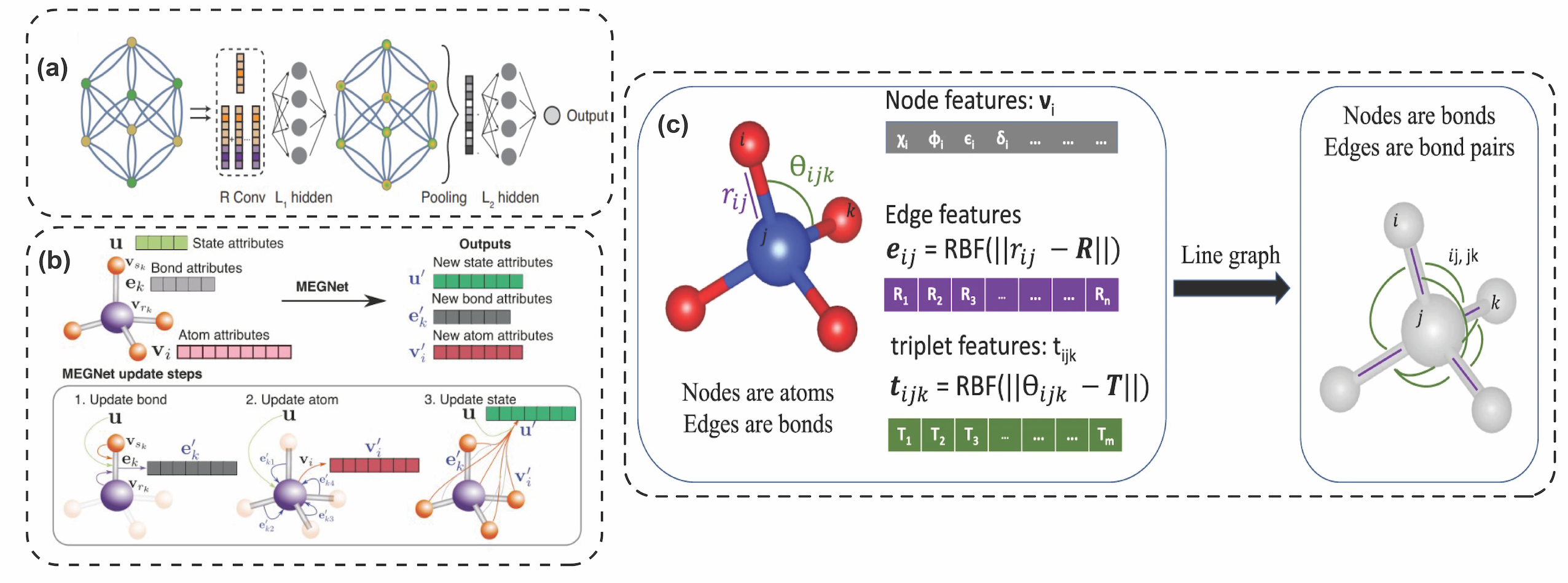}
  \caption{\label{fig:S2} (a) Crystal graph convolutional neural network (CGCNN) (b) Materials graph network (MEGNET) and (c) Atomsitic line graph neural network (ALIGNN) architecture.}
  \vspace{0.5cm}
\end{figure*}

\begin{figure*}[htb]
  \centering
\includegraphics[width=.9\linewidth]{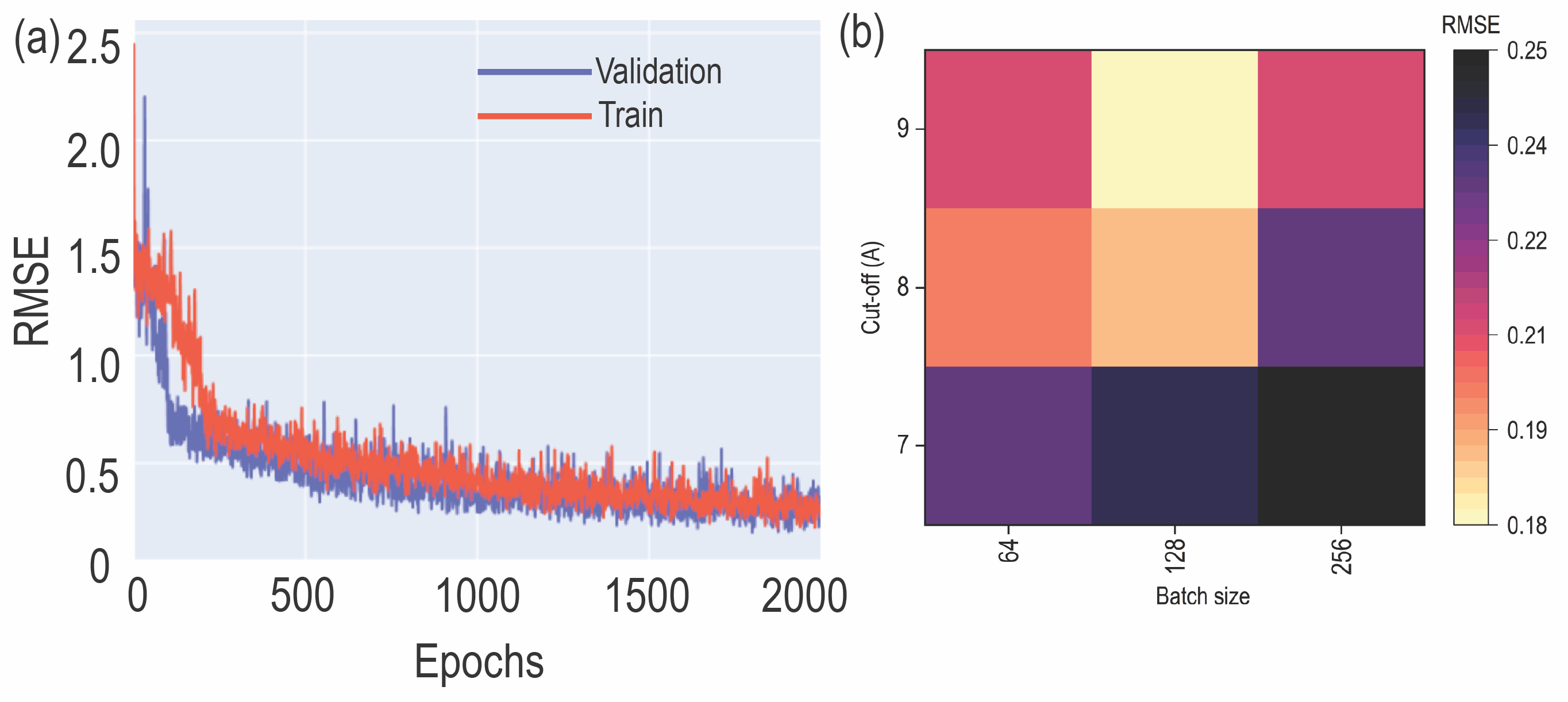}
\caption{\label{fig:S3} (a) Learning curve for a CGCNN model trained on dataset 1 under A-rich chemical potential conditions, and (b) optimization of batch size and cut-off radius in the CGCNN model.}
\vspace{0.5cm}
\end{figure*}

\begin{figure*}[htb]
  \centering
\includegraphics[width=.9\linewidth]{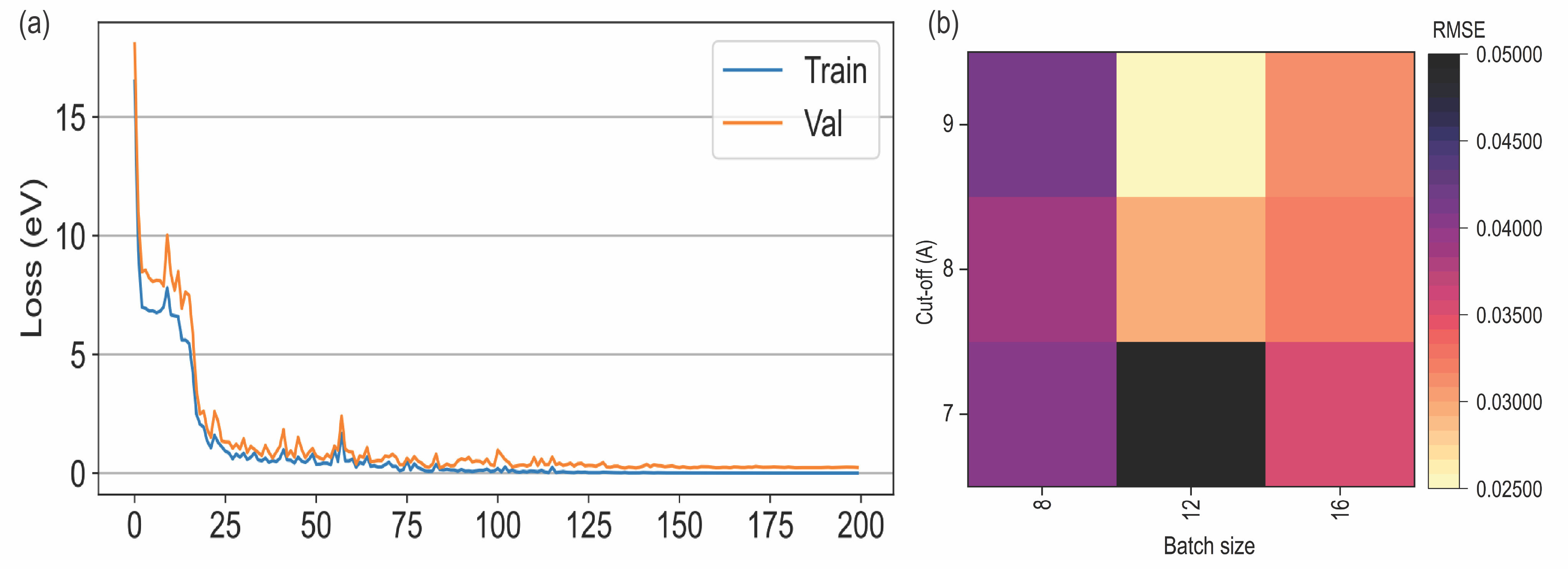}
\caption{\label{fig:S4} (a) Learning curve for an ALIGNN model trained on dataset 1 under A-rich chemical potential conditions, and (b) optimization of batch size and cut-off radius in the ALIGNN model.}
\vspace{0.5cm}
\end{figure*}

\begin{figure*}[htb]
  \centering
\includegraphics[width=.9\linewidth]{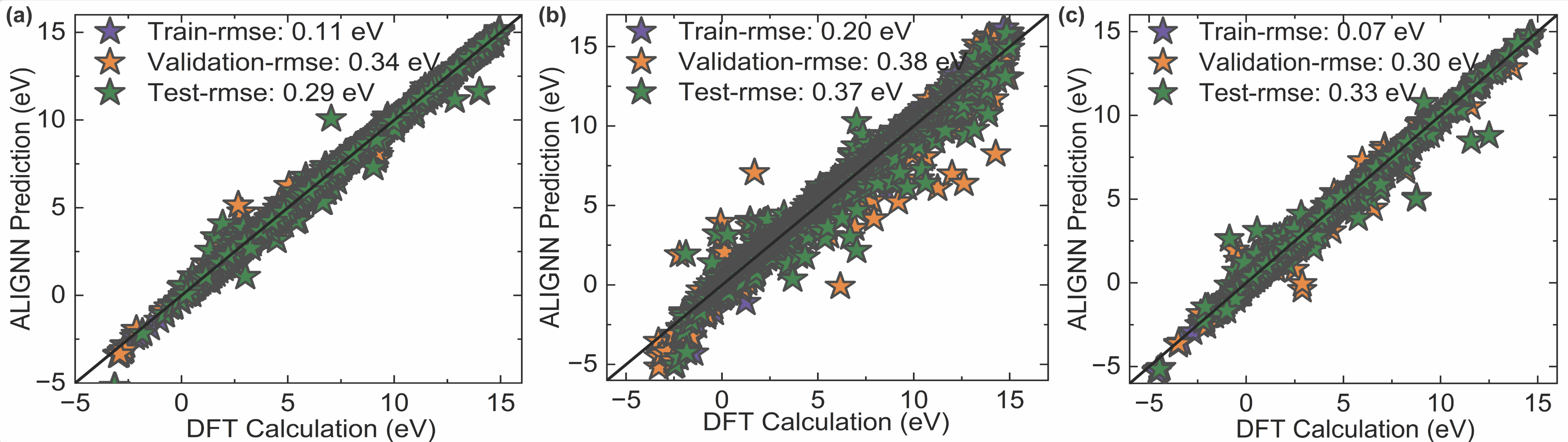}
\caption{\label{fig:S5} Parity plots for ALIGNN models corresponding to (a) E$^{f}$(q=+1, E$_{F}$=0) (b) E$^{f}$(q=0, E$_{F}$=0) (c) E$^{f}$(q=-1, E$_{F}$=0), under A-rich chemical potential conditions, prior to outlier removal.}
\vspace{0.5cm}
\end{figure*}

\begin{figure*}[htb]
  \centering
\includegraphics[width=.9\linewidth]{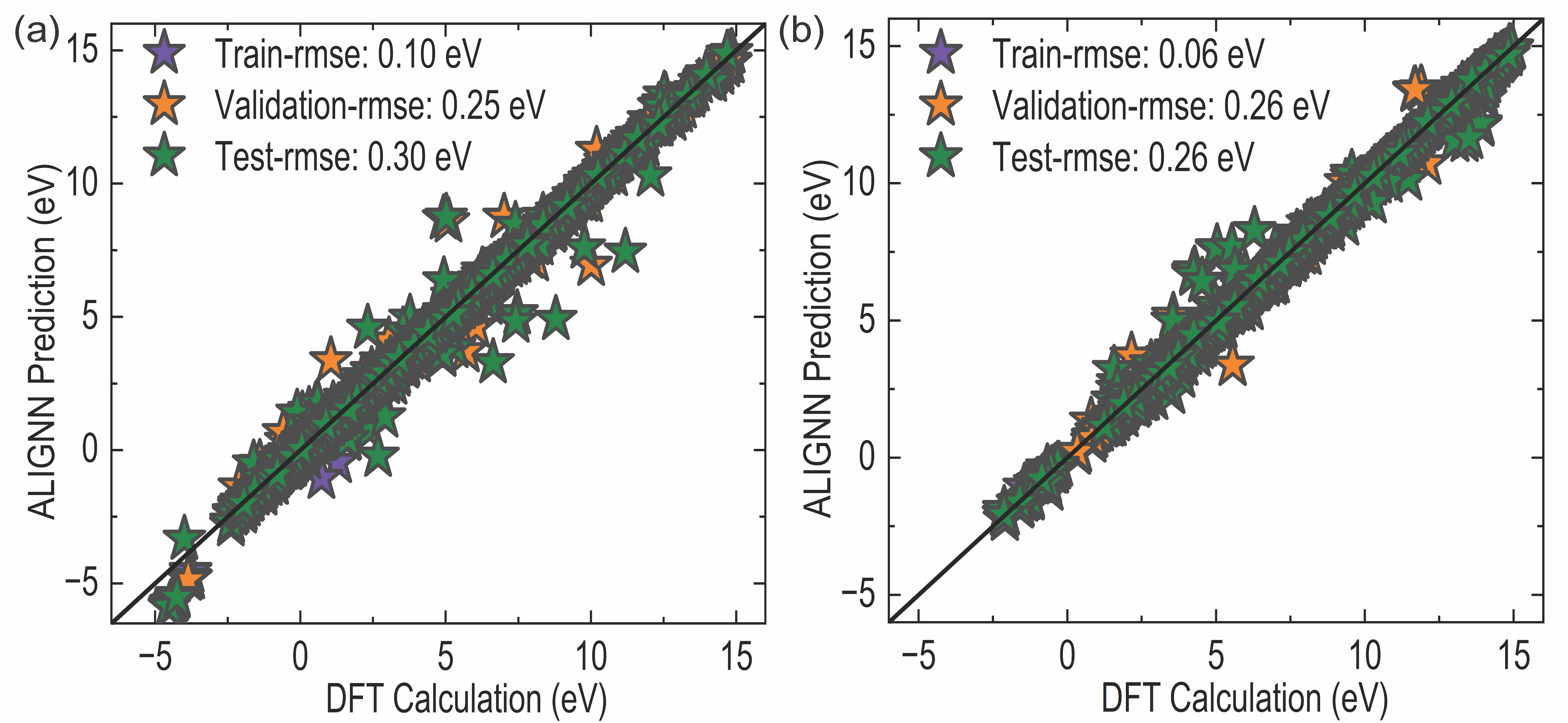}
\caption{\label{fig:S6} Parity plots for ALIGNN models corresponding to (a) E$^{f}$(q=+2, E$_{F}$=0) (b) E$^{f}$(q=-2, E$_{F}$=0), under A-rich chemical potential conditions.}
\vspace{0.5cm}
\end{figure*}

\begin{figure*}[htb]
  \centering
\includegraphics[width=.9\linewidth]{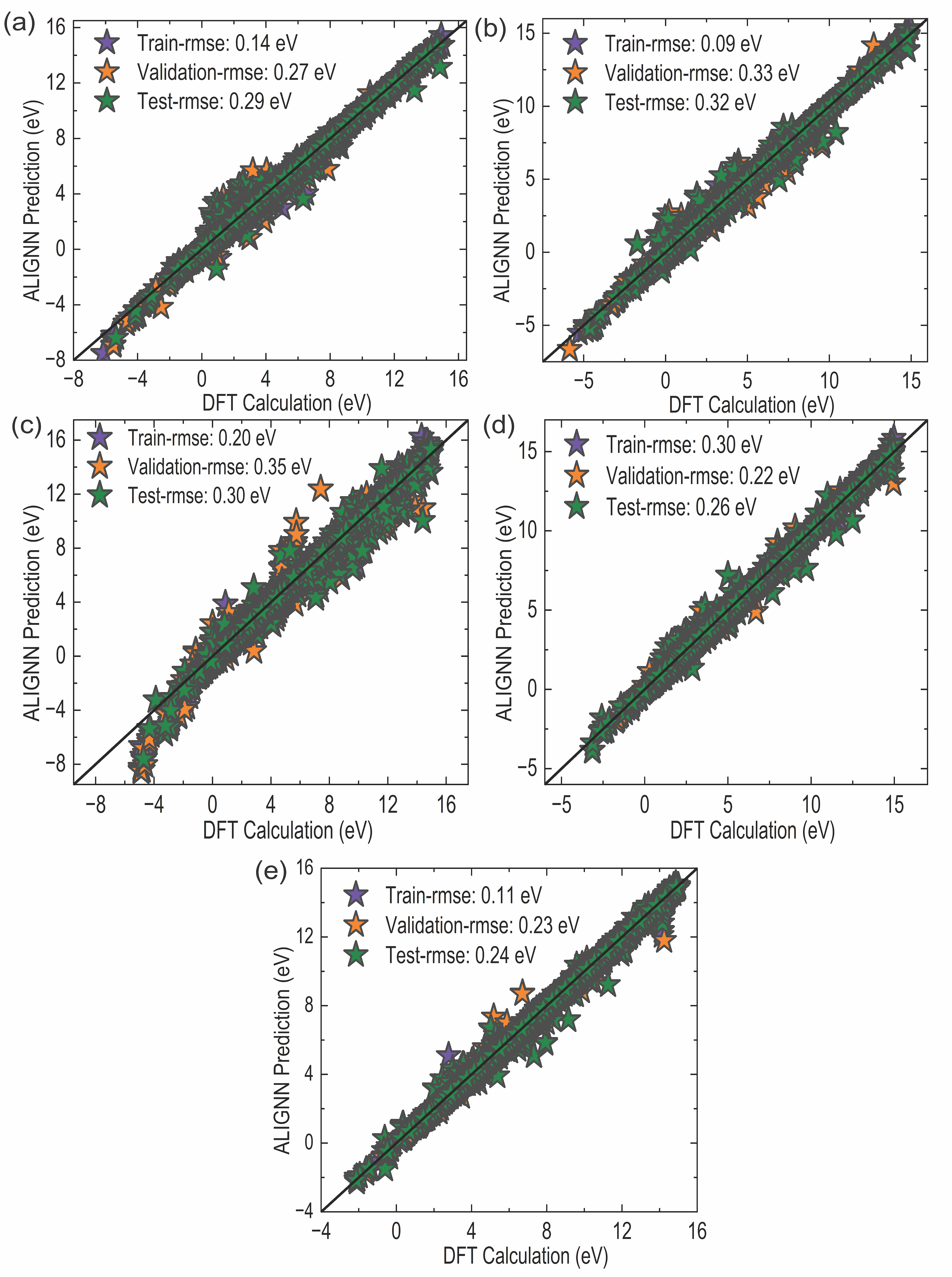}
\caption{\label{fig:S7} Parity plots for ALIGNN models corresponding to (a) E$^{f}$(q=+2, E$_{F}$=0), (b) E$^{f}$(q=+1, E$_{F}$=0), (c) E$^{f}$(q=0, E$_{F}$=0), (d) E$^{f}$(q=-1, E$_{F}$=0), (e) E$^{f}$(q=-2, E$_{F}$=0), under B-rich chemical potential conditions.}
\vspace{0.5cm}
\end{figure*}

\begin{figure*}[htb]
  \centering
\includegraphics[width=.9\linewidth]{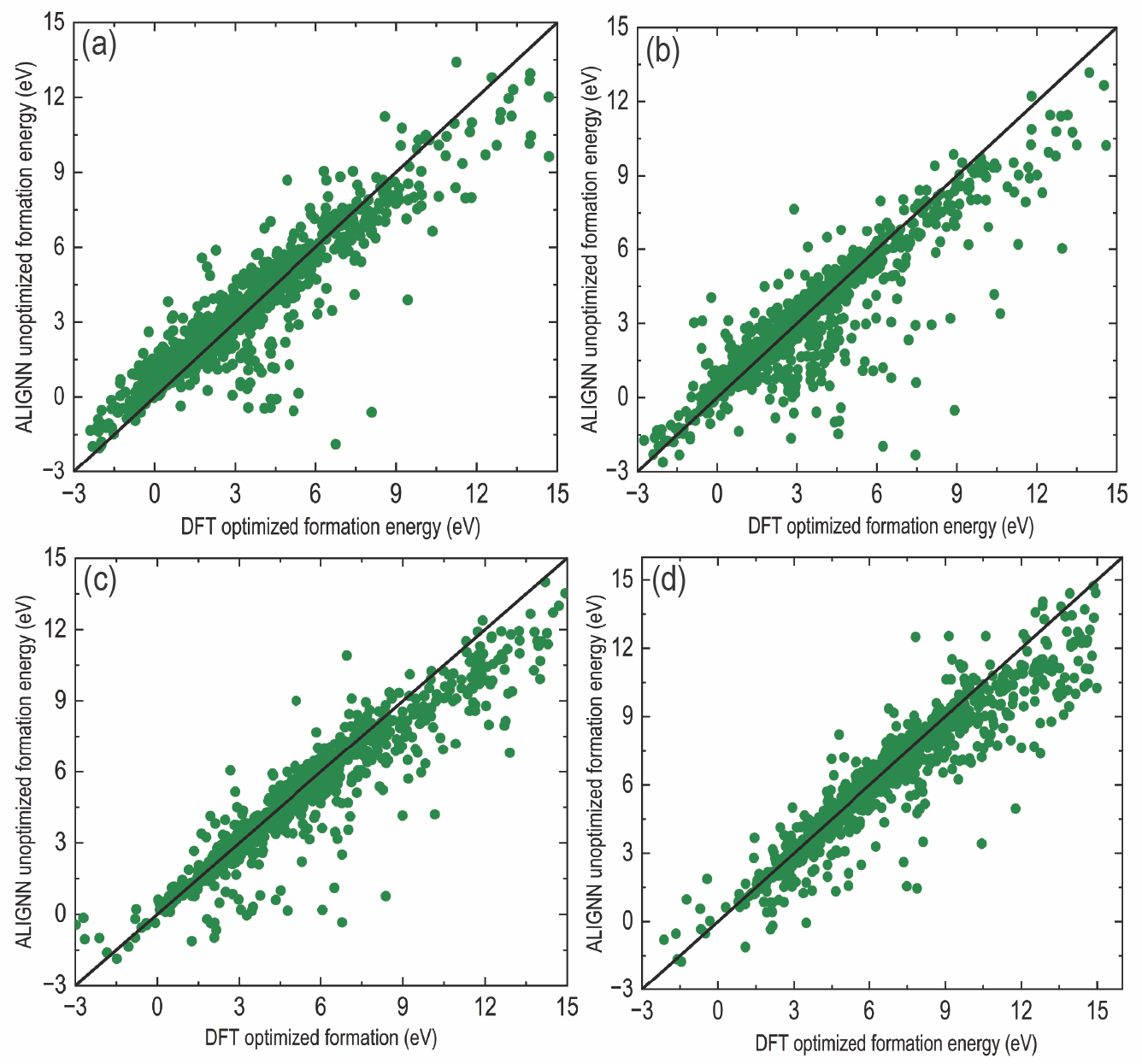}
\caption{\label{fig:S8} ALIGNN-unoptimized vs DFT-optimized (a) E$^{f}$(q=+2, E$_{F}$=0), (b) E$^{f}$(q=+1, E$_{F}$=0), (c) E$^{f}$(q=-1, E$_{F}$=0), and (d) E$^{f}$(q=-2, E$_{F}$=0), under A-rich chemical potential conditions.}
\vspace{0.5cm}
\end{figure*}

\begin{figure*}[htb]
  \centering
\includegraphics[width=.9\linewidth]{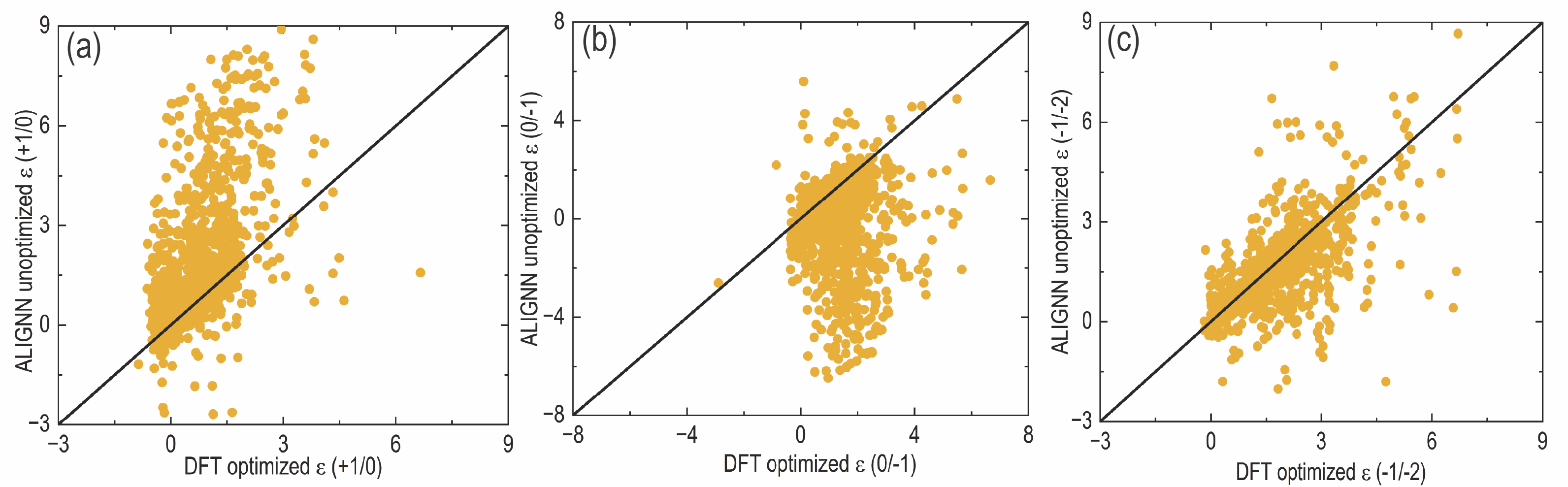}
\caption{\label{fig:S9} ALIGNN-unoptimized vs DFT-optimized (a) $\epsilon$(+1/0), (b) $\epsilon$(0/-1), and (c) $\epsilon$(-1/-2) charge transition levels.}
\vspace{0.5cm}
\end{figure*}

\begin{figure*}[htb]
  \centering
\includegraphics[width=.9\linewidth]{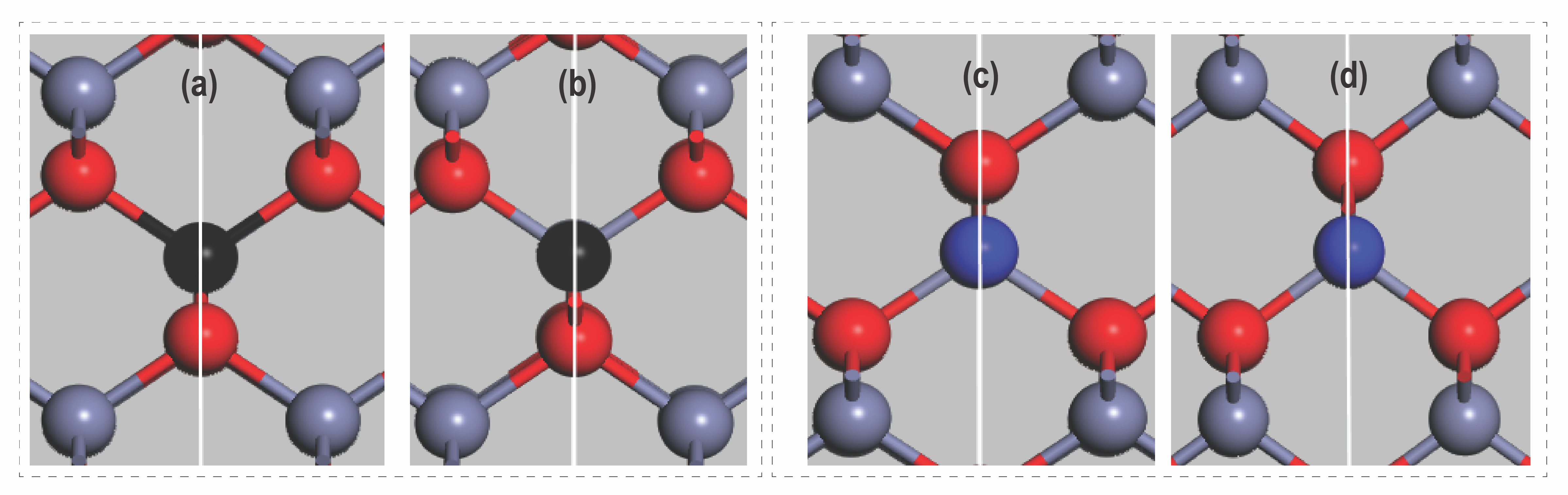}
\caption{\label{fig:S10} Atomic structure snapshot of Re\textsubscript{Zn} in ZnO, (a) after full DFT optimization, and (b) after ALIGNN optimization. Atomic structure snapshot of La\textsubscript{Zn} in ZnO (c) after full DFT optimization, and (d) after ALIGNN optimization. Grey, red, black, and blue depict Zn, O, Re, and La, respectively.}
\vspace{0.5cm}
\end{figure*}

\end{document}